\documentclass[11pt]{article}

\usepackage{amsmath}
\usepackage{enumerate}
\usepackage{amsfonts}
\usepackage[all]{xy}
\usepackage[colorlinks=true,linktocpage=true,linkcolor=blue,citecolor=blue]{hyperref}
\usepackage[numbers,sort&compress]{natbib}
\usepackage{graphicx}
\usepackage{amsmath,graphicx,amssymb,subfigure}
\usepackage{amssymb}
\usepackage[section]{placeins}

\newcommand{\bea}{\begin{eqnarray}}
\newcommand{\eea}{\end{eqnarray}}

\def\nc {N_\textrm{c}}

\def\ft#1#2{{\textstyle{\frac{\scriptstyle #1}{\scriptstyle #2} } }}
\def\fft#1#2{{\frac{#1}{#2}}}

\newcommand{\cf}{{\cal F}}
\newcommand{\cb}{{\cal B}}
\newcommand{\ch}{{\cal H}}

\newcommand{\uh}{u_\textrm{H}}
\newcommand{\g}{\gamma}
\newcommand{\cs}{c_\textrm{sch}}

\begin{document}

\bibliographystyle{hieeetr}

\pagestyle{plain}
\setcounter{page}{1}

\begin{titlepage}

\begin{center}

\vskip 12mm

{\LARGE {\bf Anisotropic plasma at finite $U(1)$ chemical potential}}

\vskip 1 cm

{\large {\bf Long Cheng$^{1}$ ,  Xian-Hui Ge$^{1}$, Sang-Jin Sin$^{2}$}}\\

\vskip .8cm

{\it $^{1}$ Department of Physics, Shanghai University\\
Shanghai 200444, P.R China}\\
{\it $^{2}$ Department of Physics, Hanyang University\\
Seoul 133-791, Korea}\\

\medskip

\vskip 0.5cm

%{\tt  physcheng@shu.edu.cn, \, gexh@shu.edu.cn,}

\vspace{5mm}
\vspace{5mm}

{\bf Abstract}\\
\end{center}
We present a type IIB supergravity solution dual to a spatially anisotropic  $\mathcal{N}=4$ super Yang-Mills plasma at finite $U(1)$ chemical potential and finite temperature. The effective five-dimensional gravitational theory
is a consistent Einstein-Maxwell-dilaton-Axion truncation of the gauged supergravity. We obtain the solutions both numerically and analytically. We study the phase structure and thermodynamic instabilities of the solution, and find new instabilities independent of the renormalization scheme.

 \noindent

\end{titlepage}

\tableofcontents

%%%%%%%%%%%%%%%%%%%%%%%%%%%
\setlength\arraycolsep{1pt}
\section{Introduction}

 The quark-gluon plasma (QGP) created at the Relativistic Heavy Ion Colliers (RHIC) \cite{rhic,rhic2} and at the Large Hadron Collider (LHC)  shows its strongly coupled fluid behavior \cite{fluid,fluid2}, and thus its physical explanation calls for non-perturbative computational methods. A powerful tool for studying the strongly coupled  plasma is the gauge/string duality, in which the best understood example is the AdS/CFT correspondence. This correspondence asserts the equivalence between type IIB superstring theory in $AdS_5\times S^5$ and $\mathcal N=4$ Super Yang-Mills (SYM) gauge theory on the 4-dimensional boundary of $AdS_5$   \cite{duality,duality2,duality3}. By using the AdS/CFT duality, one can study many physical quantities of the dual QCD  in the strongly coupled regime (see \cite{review1,review2} for reviews).

 Another significant observation of the RHIC and the LHC experiments is that the plasma created is anisotropic and non-equilibrium during the period of time $\tau_{out}$ after the collision. That is to say, it is locally anisotropic at the time $\tau_{out}<\tau<\tau_{iso}$, a configuration to be described by the hydrodynamics with the anisotropic energy-momentum tensor \cite{ani,ani2,ani3,ani4,ani5,ani6,ani7,ani8,ani9,Kirsch1}. However, the anisotropic plasma yields its instabilities: the instabilities found in the weakly coupled regime are responsible for the isotropization \cite{Weibel:1959zz,Mrowczynski:1988dz,Mrowczynski:1993qm,Mrowczynski:1994xv,Mrowczynski:1996vh,Randrup:2003cw,Romatschke:2003ms,Arnold:2003rq,Romatschke:2004jh,Arnold:2004ti,Rebhan:2004ur,Arnold:2005vb}.
 Such instabilities in the strong coupling regime have been investigated under the framework of the AdS/CFT correspondence \cite{Janik,Steineder1,Steineder10}, where the anisotropic dual geometry with a naked singularity was involved.

 More recently, a remarkable progress has been achieved in the study of anisotropic gauge/string duality by Mateos and Trancanelli, who constructed a static and regular anisotropic black brane solution of the type IIB supergravity  which can dually describe a spatially anisotropic $\mathcal N = 4$  SYM plasma \cite{mateos,mateos2}. Moreover, they also revealed that, in some regions, the homogeneous phase of strong coupling plasma displays  instabilities reminiscent of weakly coupled plasmas.

Inspired by this seminal work, many studies have been focused on the effects of the anisotropy on more physical observables of the anisotropic plasma, such as  the drag force experienced by an infinitely massive quark propagating at constant velocity in this anisotropic background \cite{chernicoff}, the jet quenching parameter of this anisotropic plasma \cite{giataganas,chernicoff2,Rebhan:2012bw} and the anisotropy effect on heavy-quark energy loss \cite{energy loss}. Remarkably, the shear viscosity longitudinal to the direction of anisotropy is found violating the holographic viscosity bound \cite{Violation} which presented the first example of such violation  non-involving higher-derivative theories of gravity. More relevant studies can also be found in \cite{ad1,ad2,ad3,ad4,ad5,ad6,ad7,Chakraborty:2012dt,ad8,ad9,Jahnke:2013rca,xhg,xhg1}.

In this paper, we are going to study the instabilities of the anisotropic plasma with a finite chemical potential  through introducing the $U(1)$ charge to the black brane in the gravity dual.  One motivation for our work comes  from the fact that  the QGP may carry a non-zero chemical potential. A simply way to introduce a $U(1)$ chemical potential  is to consider a charged black brane in the gravity dual \cite{dts1,maeda2,pbab,Kundu,fll,thermalization}. Thus, as an extension, we will construct a duality between the anisotropic charged black brane and the anisotropic SYM plasma by following \cite{mateos} \footnote{ Another way to holographic study of QCD at finite chemical potential (baryon density) has been developed by considering $N_f$ D7-brane in the background of $N_c$ black D3-brane \cite{sin,sin1,myers,myers1,Karch,sin2,cotrone1,cotrone2}.}.

 In the context of  the AdS/CFT correspondence, one can introduce the $U(1)$ gauge symmetry by various Kaluza-Klein  compactifications of $D=10$ supergravity theory on $S^5$  \cite{lvh1,lvh2,Cvetic1}. Turning on the angular momenta along $S^5$, one can obtain the rotating D3-brane solutions, and the angular momenta can be identified with $U(1)^3$ charges after $S^5$ reduction, where the isometry group of $S^5$  corresponds to R-symmetry of the dual SYM theory. A special case is that all charges are equal, in which  the solution is simply the Reissner-Nordstrom-Anti de Sitter (RN-AdS) black hole \cite{van}. Following these  procedures, we will introduce the $U(1)$ gauge field to the anisotropic system  and then obtain the anisotropic charged AdS black brane solution. So to some extent, this work can be treated as
an extension of \cite{van} to the  anisotropic case. In Fig.1 we explicitly display the relations between the anisotropic and isotropic
solutions with or without U(1) gauge field.
   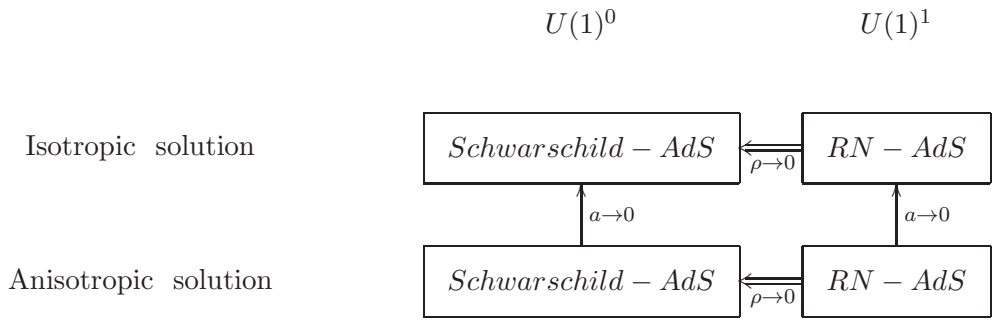
\begin{figure}[ht]
\[\xymatrix{&    &   U(1)^{0}   &   U(1)^{1}  &   \\
   {\rm Isotropic~~solution} &  & *+++[F]{Schwarschild-AdS}  & *+++[F]{RN-AdS} \ar@{=>}[l]^{~~~~~~~\rho\to0} & \\
   {\rm Anisotropic~~solution}  &   & *+++[F]{Schwarschild-AdS} \ar[u]_{a\to0} & *+++[F]{RN-AdS} \ar[u]_{a\to
   0}  \ar@{=>}[l]^{~~~~~~~\rho\to0} & }
\]
\caption{Diagram shows the relations between anisotropic and isotropic black brane solutions.  The diagram to the left up to $U(1)^{0}$ corresponds to Schwarzschild-AdS solution}. \label{adsdiagram}
\end{figure}

   It is well known that the $U(1)$ symmetry plays an important role in many condensed matter systems, which is dual to the charged black branes in AdS space. So another motivation of this paper is the future application  of the anisotropic duality with a $U(1)$ chemical potential to condensed matter systems such as liquid crystals and anisotropic Fermi surfaces. We will provide a RN-AdS version of the anisotropic black brane solution as the first step to study the anisotropic condensed matter systems.

The organization of the contents is as follows. In section 2, we  present a ``prolate"  black brane solution in the presence of the $U(1)$ field from the type IIB supergravity
 and discuss the entropy density with different charges. We show later that the zero temperature limit
  of the black brane cannot be reached unless the anisotropic parameter takes imaginary values. Meanwhile, the entropy density does not decrease in the low temperature limit just as the
  RN-AdS solution. In section 3, we study the stress tensor and thermodynamics of this prolate systems  through the holographic renormalization of the Einstein-Maxwell-Dilaton-Axion theory.  In section 4, we  discuss the thermodynamic stabilities of the prolate black brane solution and its phase structure by comparing with those of  isotropic RN-AdS black brane and uncharged anisotropic black brane, respectively. Conclusions and possible directions for future investigations are presented in section 5.

In the appendices, we present the detailed computations. In Appendix A, we  derive the equations of motion and discuss the numerical solution of the metric with the anisotropic parameter $a^2>0$. In Appendix B, we give in detail the holographic renormalization of the bulk theory. In Appendix C, we present the high temperature black brane solution and various quantities in the small charge density and weak anisotropy limits up to the $\mathcal{O}(a^4)$ order analytically. In Appendix D, we compute analytically the weak anisotropic solution by perturbating around  the isotropic RN-AdS solution.  Finally, we show the numerical and analytic solutions with $a^2<0$ which corresponds to the ``oblate" solutions in appendix E.

\section{Anisotropic charged black brane solution}
In this section, we investigate the five-dimensional Einstein-Maxwell-Dilaton-Axion  truncation of gauge AdS supergravity with  compactification of ten-dimensional type IIB supergravity on $S^5$. We aim to obtain its general anisotropic solutions, which can be contributed to exploring the duals of QCD-like gauge theories.
\subsection{Action and solution}
\label{action-section}
As anticipated in the introduction, we are interested in the solution with the form of ${\cal M} \times S^5$, where the manifold ${\cal M}$ is the solution of five dimensional supergravity with negative cosmological constant
$\Lambda = -6/L^2$. Note that the AdS radius $L$ measures the size of $S^5$ which is set as $L=1$ in the following discussions. To construct the solutions one need, one can reduce the ten-dimensional  supergravity action on  five-dimensional spherical internal space $S^5$ by utilizing the non-linear Kaluza-Klein reduction
ans\"atze developed in \cite{lvh1,lvh2}.
So, let us start with bosonic sector of the type IIB supergravity action $S=\frac{1}{2 \kappa_{10}^2}\int {\cal L}$ in the Einstein frame ,
\bea
{\cal L}= \hat{R}*1-\frac{1}{2}d\hat{\phi}\wedge*d\hat{\phi}-\frac{1}{2}e^{2\hat{\phi}} \hat{F}_1\wedge*\hat{F}_1-\frac{1}{4}\hat{F}_5\wedge*\hat{F}_5,
\label{10daction}
\eea
where we have truncated out two (NS-NS and R-R) 2-form potentials. $\hat{\phi}$ and $\hat{F}_1=d\hat{\chi}$ are the dilaton and the axion field-strength in ten-dimensions respectively.
The 5-form field $\hat{F}_5$ should satisfy the self-duality condition imposed at the level of equation of motion.

The equations of motion following from the Lagrangian above are read as\cite{lvh3}
\bea
&&\hat{R}_{MN}-\frac{1}{2}\partial_M\hat{\phi}\partial_N\hat{\phi}-\frac{e^{2\hat{\phi}}}{2}\partial_M\hat{\chi}\partial_N\hat{\chi}-\frac{1}{4\times4!}\hat{F}_{MPQRS}\hat{F_N}^{~PQRS}=0, \nonumber\\
&&d*d\hat{\phi}-e^{2\hat{\phi}}\hat{F_1}\wedge* \hat{F_1}=0,     \nonumber\\
&&d(e^{2\hat{\phi}}*\hat{F_1})=0, \nonumber\\
&&d(*\hat{F_5})=0, \label{10deq}
\eea
with self-duality condition
\bea
*\hat{F_5}=\hat{F_5}.
\eea
Hereafter, we use the uppercase Latin alphabet and lowercase Greek alphabet for ten and five dimensional bulk indices respectively (i.e. $M,N,..=0,..,9$, $\mu, \nu=0,..,4$). The boundary indices are represented by lowercase Latin alphabet  as usual (i.e. $i,j=0,..,3$).

Since we only want to reduce the theory into minimal supergravity, the reduction ansatz is given by
\begin{eqnarray}
ds_{10}^2 &=& ds_5^2 + \Big[d\xi^2 + s^2 (d\tau - \frac{1}{ \sqrt3} A_1)^2 + \frac{1}{4} c^2 \Big(\sigma_1^2 + \sigma_2^2 + (\sigma_3 - \ft2{\sqrt3} A_1)^2\Big)\Big],\nonumber\\
\hat{F}_{5} &=& H_5 + {* H_5},\nonumber \\
H_{5} &=& 4 \epsilon_5 + \fft{1}{\sqrt3}\,{*_5 F_2}\, \wedge \Big(\ft18 c^2 \sigma_1\wedge\sigma_2\nonumber\\&&~~~~~+\ft12(\sigma_3 - \ft2{\sqrt3}A_1)^2\wedge d\xi + sc (d\tau - \ft1{\sqrt3} A_1) \wedge d\xi\Big),\
\end{eqnarray}
where $*_5$ is the five-dimensional Hodge operator, $\epsilon_5$ is the volume form on the reduced five-dimensional space,  the 2-form $F_2=dA_1$ and $\sigma_i$ are the $SL(2,R)$ left-invariant 1-forms. Besides, we introduce the ansatz for $SL(2,R)$-coset scalar $(\hat{\phi}, \hat{\chi})$
\bea
\hat \phi(x,y) = \phi (x)\,,\qquad \hat \chi(x,y) = \chi(x),\,
\eea
where   $x$ denotes  the coordinates of the lower-dimensional space-time while $y$ is the ``compactifying" dimensional coordinate.

Substituting these ans¡§atz into the equations of motion (\ref{10deq}) for the type IIB theory, we can obtain the five-dimensional equations of motion that can be derived from the following Lagrangian
 \bea
 {\cal L}=\sqrt{-g} \Big( R +12-\ft12(\partial\phi)^2 - \ft12 e^{2\phi} (\partial \chi)^2- \ft14 F_2^2\Big) + \ft{1}{12\sqrt3} \epsilon^{\mu\nu\rho\sigma\lambda} F_{\mu\nu} F_{\rho\sigma} A_\lambda\,.
 \label{lag}
\eea
For simplicity, we  consider the static electrically charged solution, which means that
\bea
A&=& A_t dt,
\eea
and the Chern-Simons term $\epsilon^{\mu\nu\rho\sigma\lambda} F_{\mu\nu} F_{\rho\sigma} A_\lambda$  vanishes. So the five-dimensional effective
action we need here is given by
\bea
S&=&\frac{1}{2 \kappa^2}\int \, (R+12)*_51-\frac{1}{2}d\phi\wedge*_5d\phi-\frac{1}{2}e^{2\phi} F_1\wedge*_5F_1-\frac{1}{2}F_2\wedge*_5F_2, \label{5d_action}
\eea
where $2\kappa^2 = 16 \pi G_5$ is the five-dimensional gravitational coupling and $\kappa^2=4\pi^2/\nc^2$ is followed from $L=1$.  Then the equations of motion for the dilaton, the $U(1)$ gauge field and the gravitational field are given by
 \bea
&&\nabla_{\mu}\nabla^{\mu}\phi-e^{2\phi}(\partial\chi)^2=0\,,
\label{dilatonEOM}\\
&&\nabla_{\mu}F^{\mu\nu}=0,
\label{MaxwellEOM}\\
&&R_{\mu\nu}-\frac{1}{2}\partial_{\mu}\phi\partial_{\nu}\phi-\frac{e^{2\phi}}{2}\partial_{\mu}\chi\partial_{\nu}\chi-\frac{1}{2}F_{\mu\lambda}F_\nu^{~\lambda}+\frac{g_{\mu \nu}}{12}F_{\lambda\rho}F^{\lambda\rho}+4g_{\mu\nu}=0~~.
 \label{EinsteinEOM}
\eea
 Following the anisotropic ansatz  in \cite{mateos}, we assume that the metric takes the form
\bea
&&\hskip -.35cm
ds^2 =  \frac{e^{-\frac{1}{2}\phi}}{u^2}\left( -\cf \cb\, dt^2+dx^2+dy^2+ \ch dz^2 +\frac{ du^2}{\cf}\right),
\,\,\,\,\,\,\label{ansatz1}
\eea
together with
\bea
  \phi=\phi(u) , ~~~~A_t=A_t(u) \,,
\label{sol2}
\eea
where the axion field is set to be linear in the asymmetric direction as $\chi=az$. The metric coefficients $\cf, \cb$ and $\ch$  only
depend on the holographic radial coordinate $u$. Apparently, the spatial anisotropy is due to the presence of nontrivial $\ch$. The horizon of the black brane is defined by $u=\uh$ at which $\cf(u)=0$. The asymptotical AdS boundary of the spacetime locates at $u=0$.

As shown in the appendix \ref{derivation}, under the ansatz of (\ref{ansatz1}), the Maxwell equations can be solved directly as
\bea
F=Q\sqrt{\cb}e^{\frac{3}{4}\phi}u dt\wedge du,
\eea
where $Q$ is a  constant related to the $U(1)$ charge density. The corresponding chemical potential $\mu$ is then given by
\bea
\mu=\int_0^{\uh}Q\sqrt{\cb}e^{\frac{3}{4}\phi}u du. \label{chenmecial}
\eea
Using the equations of motion, we can express the metric in terms of the dilaton
\bea
\ch&=& e^{-\phi} , \label{eq_H} \\
\cf&=&\frac{e^{-\frac{1}{2}\phi}}{12(\phi'+u\phi'')}\left(3a^2 e^{\frac{7}{2}\phi}(4u+u^2\phi')+48\phi'-2e^{\frac{5}{2}\phi}Q^2u^6\phi'\right),
\label{eq_F} \\
\frac{\cb'}{\cb}&=&\frac{1}{24+10 u\phi'}\left(24\phi'-9u\phi'^2+20u\phi''\right)\,\label{eq_B},
\eea
where the prime $'$ denotes derivative with respect to $u$. The dilaton itself should satisfy a  third-order equation
\bea
0&=&\frac{-48 \phi '^2 \left(32+7 u \phi '\right)+768 \phi ''+4 e^{\frac{5 \phi }{2}} Q^2 u^5 \left(-24 \phi '+u^2 \phi
    '^3-8 u \phi ''\right)}{48 \phi '-2 e^{\frac{5 \phi}{2}} Q^2 u^6 \phi '+3 a^2 e^{\frac{7 \phi}{2}} u \left(4+u \phi '\right)}
    +\frac{1}{u \left(12+5 u \phi '\right) \left(\phi '+u \phi ''\right)}\nonumber\\
    && \times\Big[13u^3\phi'^4+u^2\phi'^3(96+13u^2\phi'')+8u(-60\phi''+11u^2\phi''^2-12u\phi^{(3)}) \nonumber\\
    &&~~~~~+2u\phi'^2(36+53u^2\phi''-5u^3\phi^{(3)})+\phi'(30u^4\phi''^2-64u^3\phi^{(3)}-288+32u^2\phi'')\Big].
   \label{eq_dil}
\eea
The asymptotical $AdS_5$ boundary conditions require the following   constraints: $\phi(0)=0$,  $\cf(0)=\cb(0)=1$, and then $\ch(0)=1$.
Once we solve  these differential equations, we can obtain the anisotropic charged black brane solution. We  show the numerical and perturbative  solutions in the Appendix \ref{numerical}, \ref{highApp} and \ref{low App}.

One thing we should point out is that there are two classes of solutions: ``prolate" solution and ``oblate" solution,  which correspond to $\ch(\uh)>1$ and $\ch(\uh)<1$, respectively. As can be seen from Appendix \ref{numerical} and \ref{imag}, the ``prolate" solution
is obtained in the case $a^2>0$, and  the ``oblate" solution can be obtained  when $a^2<0$. In what follows in the main text, we   mainly work on the ``prolate" case with $a^2>0$ as in \cite{mateos2}, since  the real-valued constant $a$ has a clear physical explanation in the D3/D7 model \cite{takaya}. For those who are interested in the ``oblate" solution may refer to Appendix \ref{imag}. \footnote{Surprisingly, we find that the prolate and oblate solutions have very different thermodynamic properties, which   were  reported in  \cite{oblate}.}

\subsection{Temperature and entropy}
Once we  obtained the metric, the temperature and the entropy of horizon can be evaluated directly. We will see that, compared with chargeless anisotropic solution, the anisotropic charged black brane has some special thermal properties.

 After the Euclidean continuation of the metric (\ref{ansatz1}) in near horizon $\uh$ limit, the imaginary time must have periodicity $- 4\pi/\cf'(\uh)\sqrt{\cb_H }$ to avoid the conical singularity, which can determine the temperature of black brane horizon as
 \bea
T &=& -\frac{\cf'(\uh) \sqrt{\cb_\textrm{H}}}{4\pi}\nonumber\\
 &=& \sqrt{B_{H}}\bigg[\frac{e^{-\frac{\phi_H}{2}}}{16\pi \uh}(16+a^2 e^{7\frac{\phi_H}{2}}\uh^2)-\frac{e^{2\phi_H}Q^2\uh^5}{24\pi}\bigg] .
\label{temperature}
\eea
The entropy density can be simply obtained from the Bekenstein-Hawking entropy formula
\bea
s = \frac{A_\textrm{H}}{4 G V_3} =
\frac{\nc^2 e^{-\frac{5}{4}\phi_H}}{2\pi \uh^3},
\label{entropy1}
\eea
where $V_3$ is the volume of the black brane horizon in the spatial directions.

From the temperature and the entropy given above, we can obtain a remarkable thermal property for this anisotropic charged black brane. It seems that anisotropic black brane would become extremal when the charge $Q$ satisfies the relation
\bea
Q_{ext}=\sqrt{\frac{3}{2\uh^6}}\sqrt{16e^{-\frac{5}{2}\phi_H}+e^{\phi_H}a^{\frac{14}{7}}\uh^2}.\label{extremal con}
\eea
However, the extremal case cannot be realized.\footnote{Obviously, the extremal RN-AdS is a special case (isotropic) here. But for anisotropic solution, the extremal black brane  does exist in oblate solutions, see appendix \ref{imag}.}
\begin{figure}[htbp]
 \begin{minipage}{1\hsize}
\begin{center}
%\vspace*{10mm}
\includegraphics*[scale=0.80] {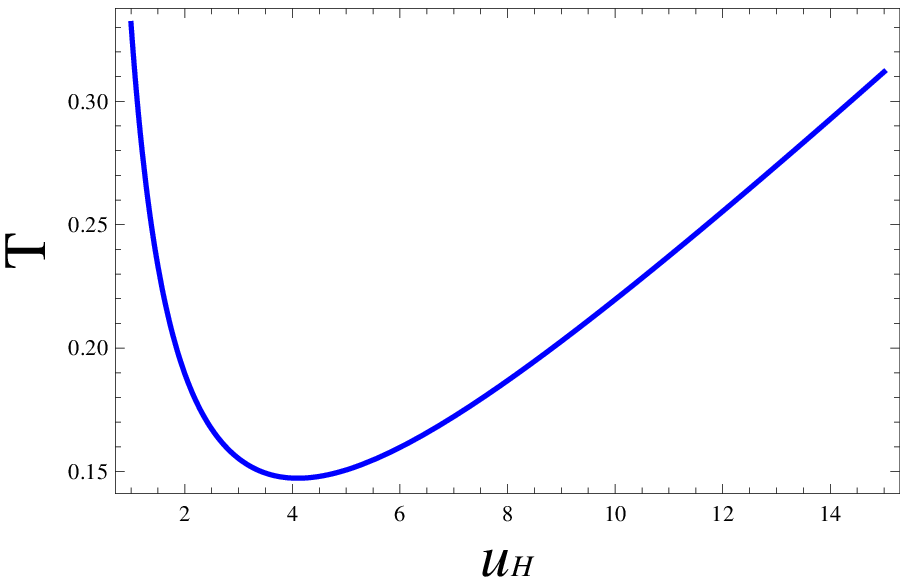}
\includegraphics*[scale=0.80]{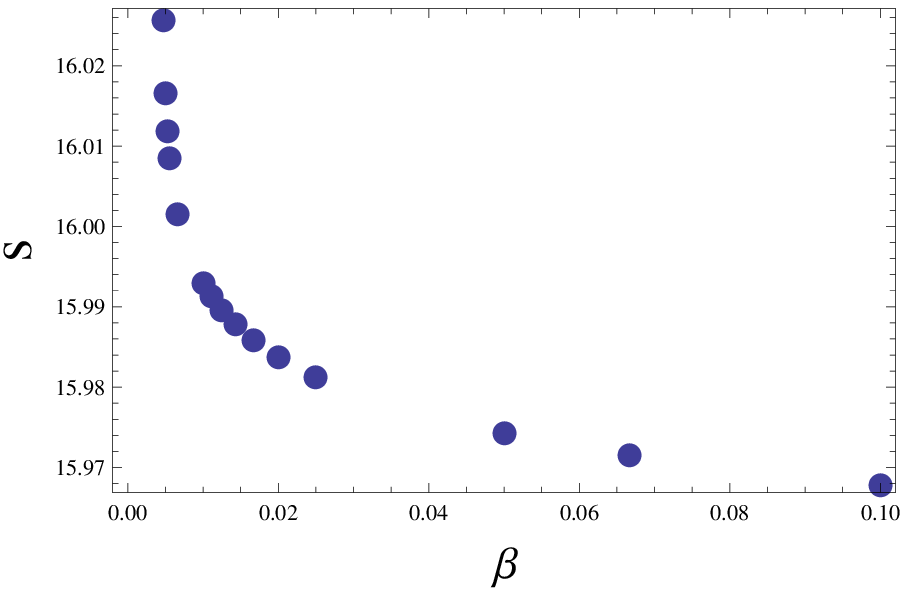}
\end{center}
\caption{(color online) The temperature as a function of the inverse horizon radii, where $a=1.46$ and $Q=0.01$. On the right, we plot the entropy density as a function of temperature $\beta\propto T$, where we fixed $\tilde\phi_H =-3/2$.}\label{beta} % where we set $\tilde\phi_H=-3/2$. }
\end{minipage}
\end{figure}
To see this explicitly, we plot the temperature as a function of the inverse horizon radius $\uh$ for fixed $a$ and $Q$ in Fig.\ref{beta} (left), which shows that the temperature has a minimal positive value. In principle, the temperature becomes zero when $\beta\equiv-\cf'(\uh)\rightarrow 0$. Nevertheless, the numerical analysis also tells us that for initial value $\tilde\phi_H \equiv\phi_H+4\log a/7=-3/2$, there will be no solution to the equations of motion in the range $\beta\leq \frac{1}{220}$  and the numerical computation breaks down there \footnote{A similar discussion for the Einstein- Maxwell-dilaton-axion SL(2,R) model can be found in\cite{maeda}.}.

 It is also of interest to checking the third law of thermodynamics which states that the area of the black brane horizon approaches zero as the temperature decreases to zero. From equation (\ref{entropy1}), we can see that for fixed $\uh$, the entropy density $s\rightarrow 0$ as $\phi_H\rightarrow \infty$. However, the numerical analysis shows that $\phi_H< 0$ for arbitrary $\beta$, implying that the entropy density is larger than that of the RN-AdS black brane and zero entropy density cannot be reached as the temperature drops (see Fig.\ref{beta}(right)). We will discuss this behavior in detail in section 4.

\begin{figure}[htbp]
 \begin{minipage}{1\hsize}
\begin{center}
%\vspace*{10mm}
\includegraphics*[scale=0.50]{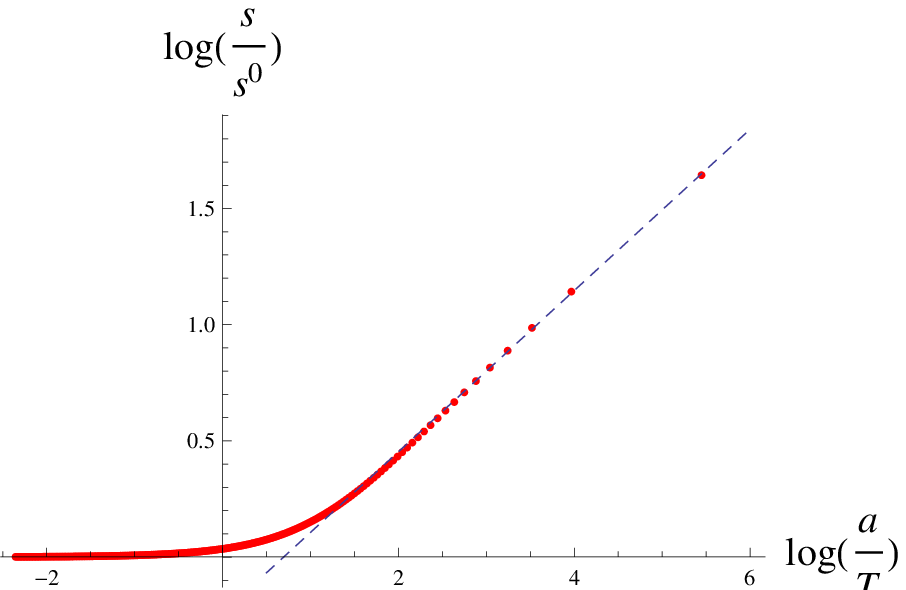}
\includegraphics*[scale=0.50]{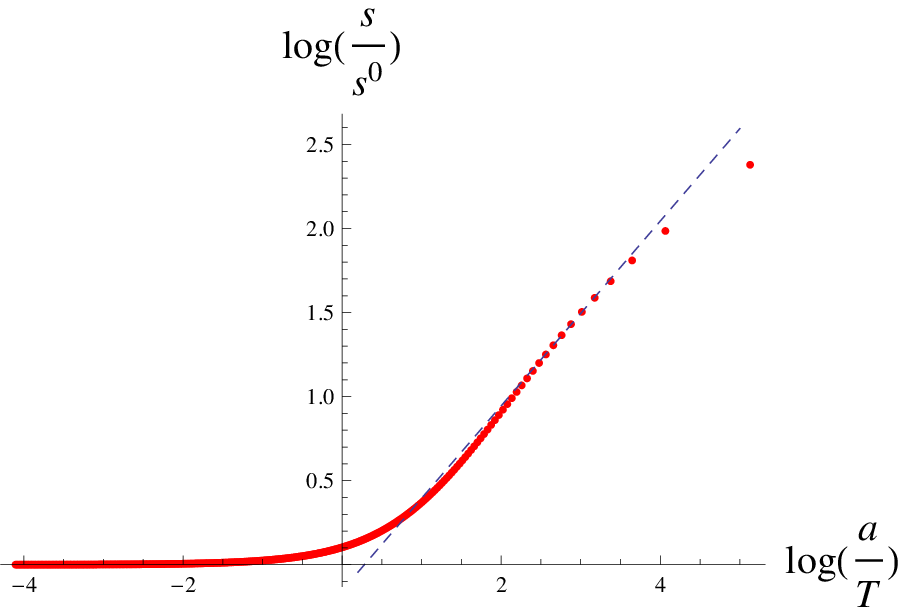}
\includegraphics*[scale=0.50]{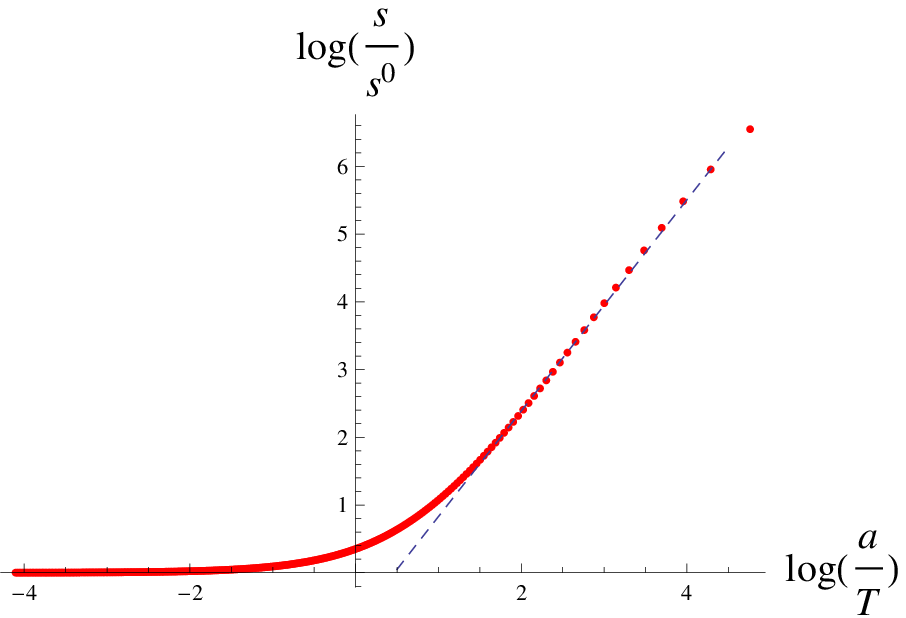}
\end{center}
\caption{(Color online.)Log-log plot of the entropy density as a function of $a/T$ for various charge  $\overline{Q}=1$(left), $\overline{Q}=2$(middle), $\overline{Q}=4$(right) where $s^0$ denotes the entropy density of RN-AdS black holes.
The dashed blue lines are straight lines with slope $0.35$ (left), $0.54$(middle) and $1.56$(right).} \label{figentropy}
\end{minipage}
\end{figure}

\begin{figure}[htbp]
 \begin{minipage}{1\hsize}
\begin{center}
%\vspace*{10mm}
\includegraphics*[scale=0.50] {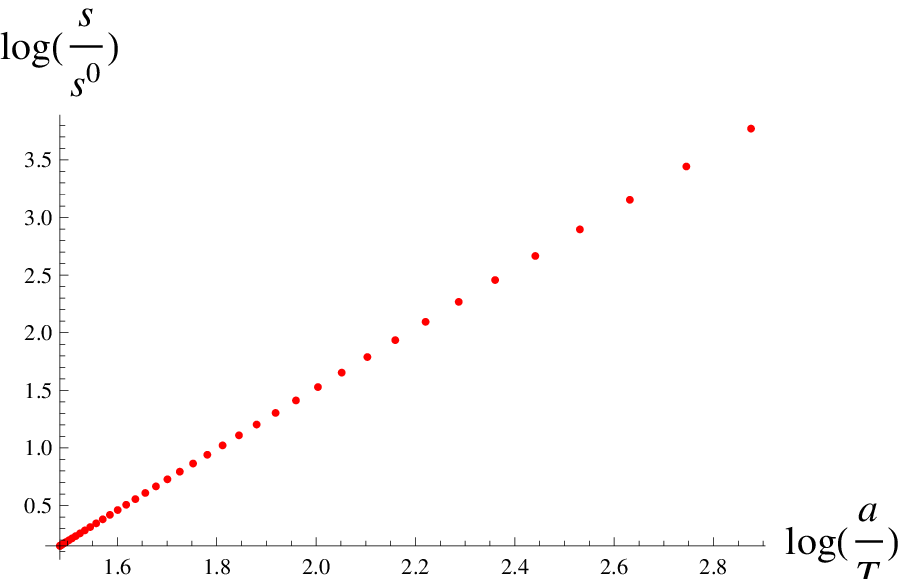}
\includegraphics*[scale=0.50] {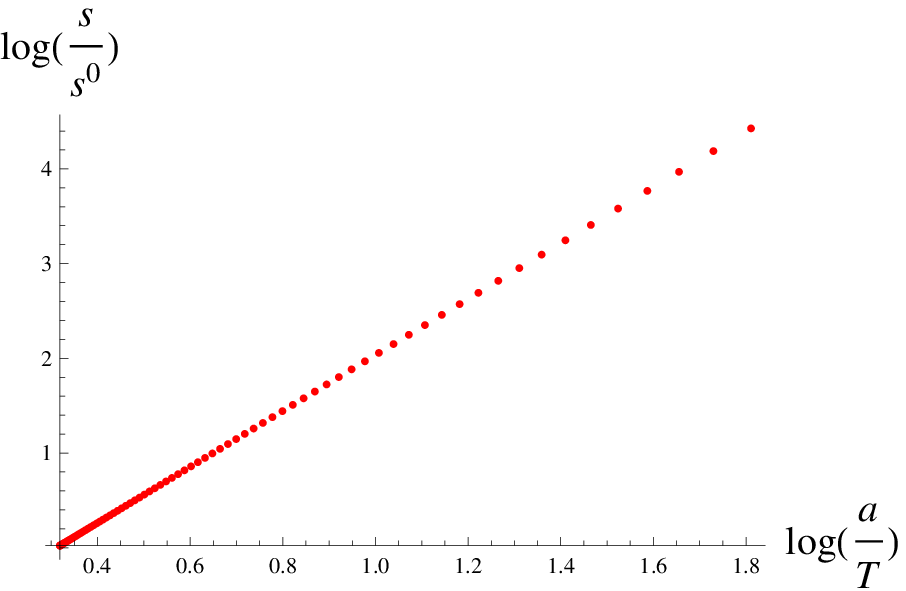}
\includegraphics*[scale=0.50]{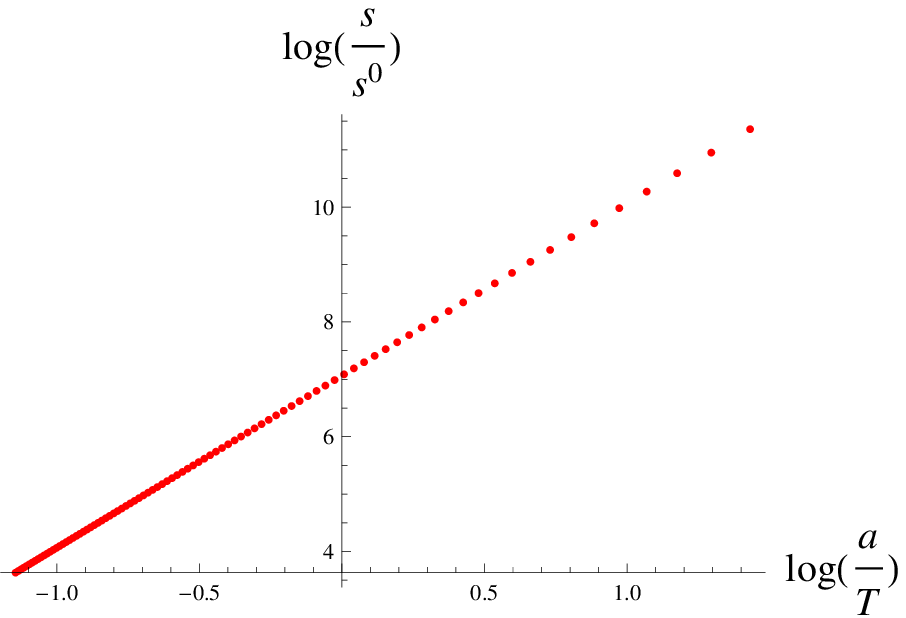}
\end{center}
\caption{(Color online.) Log-log plot of the entropy density as a function of $a/T$ for initial conditions $\tilde{\phi}_H$: $\tilde{\phi}_H=1/10$ (left), $\tilde{\phi}_H=-1/2$(middle), $\tilde{\phi}_H=-2$(right) where $s^0$ denotes the entropy density of the RN-AdS black brane. The slope from left to right is 2.93, 2.79 and 2.90.} \label{figentropy1}
\end{minipage}
\end{figure}

In order to compare the behavior of entropy density with that of \cite{mateos}, we plot the  entropy density as a function of $a/T$ for various charges $\bar{Q}\equiv a^{-5/7}Q$ and various initial conditions $\tilde{\phi}_H$. Fig. \ref{figentropy} shows the entropy density as a function of $a/T$ for different charge $\bar{Q}$. In the small anisotropy or high temperature limit $a\ll T$, the points are aligned along the horizontal axis, reproducing the case of RN-AdS black brane.  This means that in the weak anisotropy limit, the entropy density scales as the RN-AdS black brane and the horizon lies in the asymptotic AdS region which agrees well with \cite{mateos}. However, in the large anisotropy regime or low temperature limit $a\gg T$, the points are aligned along a straight line with slopes $0.35$, $0.54$ and $1.56$ corresponding to $\bar{Q}=1$, $\bar{Q}=2$ and $\bar{Q}=4$, respectively. Meanwhile, we can vary $\tilde{\phi}_H$ and plot the entropy density as a function of $a/T$ as shown in Fig.\ref{figentropy1}.
In \cite{mateos}, Mateos and Trancanelli argued that in the zero temperature limit, the entropy density scales as a IR Lifshitz-like behavior with $s\propto a^{1/3}$ and thus the horizon lies in the Lifshitz-like region. So the anisotropic black brane solution in large or small anisotropy limit can be regarded as a RG flow between the AdS geometry in the UV and a Lifshitz-like region in the IR.

Nevertheless, we   stress here that for charged black holes, the low temperature can be reached with weak anisotropy but large chemical potential $a\sim T\ll \mu$. It is well known that for RN-AdS black holes, the systems become  zero temperature with finite entropy density (i.e. horizon radius). Therefore, for our case, the black brane could be hot and cool in the AdS regime with finite chemical potential. Of course, in the small charge density limit, there may exist a RG flow to a Lifshtiz-like geometry in the IR (see Appendix C).

%%%%%%%%%%%%%%%%%%%%%%%%%%

\section{Stress Tensor and Thermodynamics}

In this section, we investigate the thermodynamic properties of the dual field on the boundary. We present the thermodynamic quantities such as energy density, pressure of our deformed ${\cal N}=4$ SYM theory via the computation of vacuum expectation value of the holographic stress tensor. The differences between isotropic thermodynamics and anisotropic thermodynamics are also discussed.
\subsection{ Holographic stress tensor}
\label{sec-thermodynamics}
 According to the holographic dictionary, the correlation functions are defined as the variation of the bulk on-shell supergravity action with respect to the boundary filed. For instance, the stress tensor is defined as one-point function corresponding to the boundary metric. Due to the integration region near the boundary of AdS which is at spatial infinity, the on-shell action will suffer the IR divergences while the dual gauge field theory will suffer from UV divergences as $u\rightarrow0$. Recalling the quantum field theory, one can perform the renormalization to deal with the UV divergence of correlation functions. Similarly, to obtain a well-defined on-shell action and finite correlation functions, we should subtract the infinities by a procedure of  holographic renormalization. We  follow the traditional method developed in \cite{anomaly,kostasgravity} rather than Hamiltonian formalism in \cite{Papadi,Yiannis}. Here we just collect the main results, the computation details can be found in Appendix \ref{Hrenormalization}.

Followed \cite{anomaly}, near the conformal boundary, the metric can be expanded in Fefferman-Graham form as
\bea \label{coord}
ds^2&=&g_{\mu\nu} dx^{\mu} dx^{\nu} =\frac{1}{r^2}\left(dr^2+h_{ij}(r)dx^i dx^j\right), \nonumber \\
h_{ij}(r)&=&h_{(0)ij}+r^2 h_{(2)ij}+r^4 h_{(4)ij}+2r^4 \tilde{h}_{(4)ij}\log r+\cdots.
\eea
Note that $h_{(0)ij}$ is the boundary metric, so we can simply set $h_{(0)ij}=\eta_{ij}$.

We can expand the dilation field, the gauge field and the axion field as the following :
\bea
\phi(r)=\phi_{(0)}+r^2\phi_{(2)}+r^4(\phi_{(4)}+2\psi_{(4)}\log r)+\cdots,  \label{diaton exp}
\eea
\bea
A_\mu(r)=A_{(0)\mu}+r^2A_{(2)\mu}+r^4( A_{(4)\mu}+2\tilde{A}_{(4)\mu}\log r)+\cdots , \label{Max exp}
\eea
\bea
\chi(r,z)=\chi_{(0)} + r^2 \chi_{(2)}+r^4 \left( \chi_{(4)} +2\tilde{\chi}_{(4)}\log r \ \right) +\cdots . \label{axion exp}
\eea
Since we work in the radial gauge in the absence of the magnetic field. Note that $A_{t(0)}$ and $A_{t(2)}$ can be reinterpreted as the $U(1)$ chemical potential and the charge density in the dual field theory respectively. The dilaton on the boundary is set to be $\phi_{(0)}=0$ by the requirement of asymptotically AdS. The axion is simply $\chi(r,z)=\chi_{(0)}=az$.

By solving the equations of motion recursively, we can obtain the coefficients
\bea
h_{(2)ij}&=&\mbox{diag}\left(\frac{a^2}{24},-\frac{a^2}{24},-\frac{a^2}{24},\frac{5a^2}{24}\right),\nonumber\\
\phi_{(2)}&=&-\frac{a^2}{4},
\eea
and
\bea
\tilde{h}_{(4)ij}&=&\mbox{diag}\left(\frac{a^4}{48},-\frac{a^4}{48},-\frac{a^4}{48},\frac{a^4}{16}\right),\nonumber\\
\psi_{(4)}&=&-\frac{a^4}{12} ,\nonumber\\
\tilde{A}_{(4)t}&=&0.
\label{dd}
\eea
 The first expression of (\ref{dd}) follows that $\tilde{h}_{(4)}$ is traceless\footnote{Traceless means $\mbox{Tr}\tilde{h}_{(4)}\equiv h_{(0)}^{ij}\tilde{h}_{(4)ij}=0$.}, while the last equation implies that no logarithmic divergences are generated by Maxwell field here. Other coefficients can not be determined completely by the asymptotic analysis, however we can obtain constraint conditions for them
\bea
\mbox{Tr} h_{(4)}=-\frac{11 a^4}{576},
\eea
and
\bea
A_{(4)t}=-\frac{A_{(2)t}a^2}{24}. \label{max exp}
\eea
The explicit form can be read off from the full solution  in Appendix \ref{highApp} and \ref{low App}. In order to find the expressions for $h_{(4)ij}$ as in \cite{mateos}, we recast  $h_{(4)ij}$ as
\bea
h_{(4)tt}&=& - \frac{23 {\mathbb{B}}_4}{28}- \frac{3 {\mathbb{F}}_4}{4} + \frac{2749 a^4}{16128}, \nonumber\\
h_{(4)xx}=h_{(4)yy}&=&- \frac{5 {\mathbb{B}}_4}{28}- \frac{ {\mathbb{F}}_4}{4} + \frac{71 a^4}{1792}, \nonumber\\
h_{(4)zz}&=&- \frac{13 {\mathbb{B}}_4}{28}- \frac{ {\mathbb{F}}_4}{4} + \frac{1163 a^4}{16128},
\eea
where $\mathbb{F}_4$ and $\mathbb{B}_4$  are the near-boundary expansion coefficients of the $ {O}(u^4)$ terms of the functions $\cf$ and $\cb$, respectively. Both of them are  $a-$ and $q-$dependent, as we can see explicitly in Appendix \ref{highApp}.

Since the Maxwell field gives no additional logarithmic divergence, the counter term  turns out to be the same as that of axion-dilaton-gravity system, that is to say,
 \begin{eqnarray}
S_{ct}=-\frac{1}{\kappa^2} \int d^4x \sqrt{\g}
\left( 3- \frac{e^{2\phi}}{8} \partial_i\chi\partial^i\chi \right)
+ \log r \int d^4x \sqrt{\g} {\cal A}-\frac{1}{4}(c_{sch}-1) \int d^4x \sqrt{\g} {\cal A},\nonumber\\
\end{eqnarray}
where ${\cal A}(\gamma_{ij},\phi,\chi,A_i)$ is the conformal anomaly, $\gamma_{ij}=h_{ij}/r^2$ is the induced metric on the boundary and $\g=-\det |\gamma_{ij}|$. Then the expectation value of the stress tensor is found to be
\bea
\langle T_{ij}\rangle = \mbox{diag}(E, P_{xy}, P_{xy}, P_z)\,,
\eea
with
\bea
E&=& \frac{\nc^2 }{2\pi^2}\left(-\frac{3}{4}{\mathbb{F}}_4 -
\frac{23}{28}{\mathbb{B}}_4
+\frac{2777}{16128}a^4+\frac{c_\textrm{sch}}{96}a^4\right) ,\nonumber \\
P_{xy}&=&
\frac{ \nc^2 }{2\pi^2}\left(-\frac{1}{4}{\mathbb{F}}_4 -\frac{5}{28}{\mathbb{B}}_4+ \frac{611}{16128}a^4-\frac{c_\textrm{sch}}{96}a^4\right) ,\nonumber \\
P_z&=&
\frac{ \nc^2 }{2\pi^2}\left(-\frac{1}{4}{\mathbb{F}}_4
-\frac{13}{28}{\mathbb{B}}_4
+\frac{2227}{16128}a^4+\frac{c_\textrm{sch}}{32}a^4\right),
\label{stress_tensor}
\eea
 where $c_\textrm{sch}$ denotes a scheme dependent parameter. The conformal anomaly is given by ${\cal A}=\langle T_{ij}\rangle =\frac{\nc^2 a^4}{48 \pi^2}$.
Note that the value of conformal anomaly is independent of the Maxwell field as we expected.

It was emphasized that the conformal anomaly plays an important role in the black brane thermodynamics \cite{mateos}. What is new in this paper is that we include the $U(1)$ chemical potential. The stress tensor under a rescaling of $a$, $T$ and $\mu$ transforms as
\bea
\langle T_{ij}(ka,kT,k\mu)\rangle=k^4\langle T_{ij}(a,T,\mu)\rangle+k^4\log k ~{\cal A} c_{ij},
\eea where $c_{ij}=\mbox{diag}(1,-1,-1,3)$. This in turn indicates the stress tensor must take the form
\bea
\langle T_{ij}(a,T,\mu)\rangle=a^4 t_{ij}\left(\frac{a}{T},\frac{a}{\mu}\right)+ \log\left(\frac{a}{\Lambda}\right)\frac{\nc^2 a^4}{48 \pi^2} c_{ij},
\eea
where $\Lambda$ is an arbitrary reference scale, a remnant of the renormalization process like the substraction point in QCD. Different choices of $\Lambda$ correspond to different choices of renormalization scheme.
Therefore, in our case, the physics depends on three dimensionless ratios $a/T$, $a/\mu$ and $a/\Lambda$.
\begin{figure}[htbp]
 \begin{minipage}{1\hsize}
\begin{center}
%\vspace*{10mm}
\includegraphics*[scale=0.6] {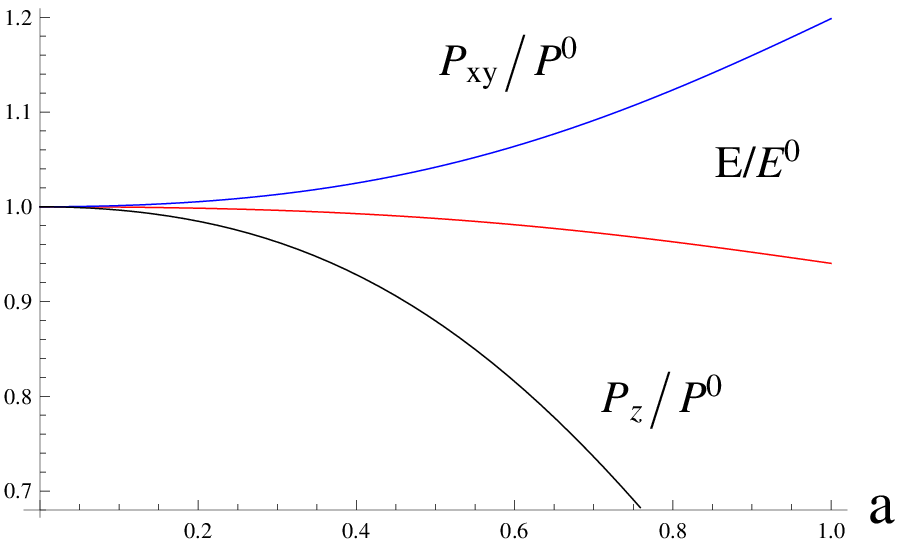}
\includegraphics*[scale=0.6] {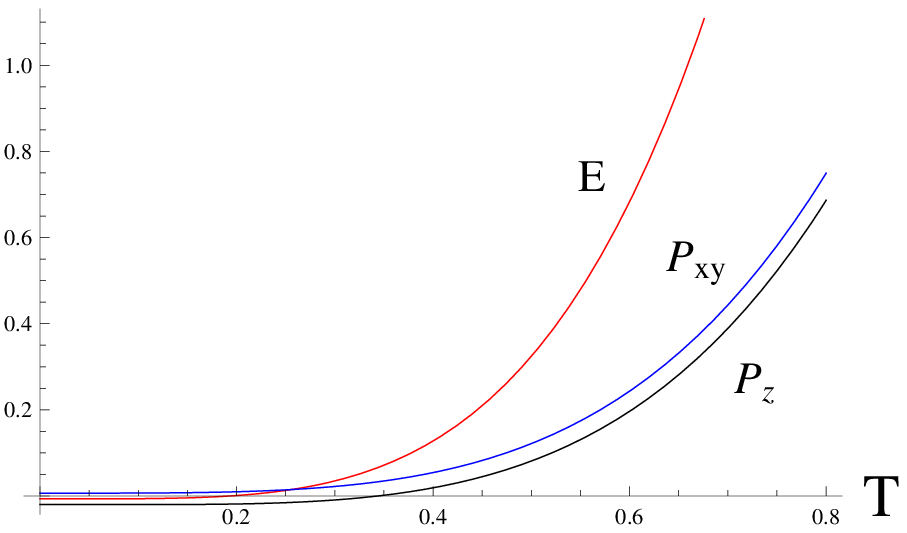}
\end{center}
\caption{(Color online.)(Left) The energy density and pressures normalized by their isotropic values as functions of $a$ with $T=0.3$ and $q=0.1$.
(Right) The energy density and pressures  as functions of $T$ with $N_c=1$, $\Lambda=10$ and $q=1/8$.}\label{figstree}
\end{minipage}
\end{figure}

The analytical expressions for the energy and the pressures (\ref{2order-stress}) and (\ref{4order-stress}) in the small anisotropy limit are obtained in Appendix C and D. Fig.\ref{figstree} shows the specific case numerically. In Fig.\ref{figstree} (left) we
plot the normalized energy and pressure for $T=0.3$ and $q=0.1$ with
\bea
E^0=\frac{6N^2_c \pi^2(1+q^2)T^4}{(2-q^2)^4},~~~~~~P^0=\frac{2N^2_c \pi^2(1+q^2)T^4}{(2-q^2)^4},
\eea
 the energy and the pressure  of the isotropic RN-AdS value, respectively.
It is clear from Fig.\ref{figstree} (left) that in the limit $a\rightarrow 0$, the energy and the pressure approach their isotropic values as expected. Fig.\ref{figstree} (right) shows that the energy or pressure could take negative value at low temperature and large $\Lambda$. We also note that in the absence of the charge density, we can recover
the anisotropic energy and pressures obtained in \cite{mateos}.

\subsection{Thermodynamics}
  As observed by the authors of \cite{mateos}, the neutral anisotropic black brane solution is interpreted to describe a uniform D7-brane charge density per unit length in the $z$-direction. The anisotropic parameter $a$ measures the number of D7-brane
 per unit length in this direction,
 \bea
 a=g_s n_{D7}=\frac{\lambda}{4\pi}\frac{n_{D7}}{N_c}.
 \label{number}
 \eea
 However, from the point of view of effective five-dimensional bulk theory, D7-branes behave as a 2-brane charge density oriented in the $xy$-directions and are homogeneously distributed along the $z$-direction. So one can introduce a chemical potential $\Phi$ conjugate to $a$. On the other hand, as shown in the  dimensional reduction, one can introduce a chemical potential  associated with the $U(1)$ global conserved charge to the framework. We thus have two ``chemical potentials'' corresponding to different charges in our anisotropic system, which have richer thermal properties than chargeless anisotropic black
brane or RN-AdS black brane.

 Let us first evaluate the free energy in the following procedure. Note that, in the saddle-point approximation, the partition function is given by
\bea
Z=e^{-S_{ren}[g*]},
\eea
where $g*$ is the Euclidean continuation of metric (\ref{sol1}) and $S_{ren}[g*]$ is the renormalized on-shell action.
Then the thermodynamic potential can be obtained by computing the Euclidean action on the analytically continued solution as
\bea
\Omega=\frac{-T\log Z}{V_3}=\frac{TS_{ren}[g*]}{V_3},
\label{Free energy}
\eea
 where the $V_3$ is the spatial volume of associated field theory. We can see later that the free energy density above is a grand canonical thermodynamic potential with respect to the $U(1)$ gauge field but canonical thermal potential with respect to axion, which means that we are working at fixed $U(1)$ potential.

 If we want to calculate the thermodynamic potential in canonical ensembles with respect to the $U(1)$ gauge field, we should add the boundary term to the on-shell action \cite{haw}.
 The   corresponding  free energy density $F$ satisfies the relation
 \bea
 F=E-Ts ,
 \eea
 and it obeys
 \bea
 dF=-sdT+\Phi d a+\mu d\rho.
 \eea
 The $U(1)$ chemical potential can be obtained from the above equation
 \bea\label{ca}
 \mu=\left( \frac{\partial F}{\partial \rho} \right)_{T,a}.
 \eea
 It is straightforward to prove that the first law of thermodynamics is satisfied
 \bea
 dE=Tds+\Phi d a+\mu d\rho,
 \eea
 where  $\rho=\frac{Q}{2\kappa^2}=\frac{\sqrt{3}q}{\kappa^2\uh^3}$ is charge density \cite{gmsst}, and $\mu$
 is the chemical potential  conjugated to $\rho$. All the black brane thermodynamic quantities are given in   Appendix C and D.

  Through the Legendre transformation, the grand canonical thermodynamic potential is related to the energy density as
 \bea
 \Omega =E-Ts-\rho\mu,
 \eea
which satisfies
 \bea\label{omega}
 d\Omega =-sdT+\Phi d a-\rho d\mu.
 \eea
 We can obtain the chemical potential conjugated to charge density $a$
\bea\label{phib}
\Phi = \left( \frac{\partial \Omega}{\partial a} \right)_{T,\mu} .
\label{sphi2}
\eea
Equation (\ref{omega}) is defined with fixed charge $a$, so it can be regarded as ``free energy" in canonical ensemble with respect to the anisotropy. However, it is the grand-canonical ensemble with respect to the $U(1)$ field, because the charge $\rho$ is free but the potential $\mu$ is fixed. Again, a consistent check can be found in
  Appendix \ref{highApp} and \ref{low App}.

The thermodynamic potential in the grand canonical ensemble with respect to $a$ is given by
\bea
G&=&E-Ts-\rho\mu-a \Phi,
\eea
and it obeys
\bea
dG=-sdT-\rho d\mu-a d\Phi.
\eea
It is straightforward to verify that
\bea
\Omega&=&-P_{xy},\\
G&=&-P_z.
\eea
So the following relation is satisfied
\bea
P_z-P_{xy}=\Phi a,
\eea
which is consistent with anisotropic fluid thermodynamics \cite{emparan}. Apparently, in the absence of the anisotropy, the relation $\Omega=G=-P$ is recovered for isotropic cases.

We should note that all the thermodynamic identities above are scheme-independent, so they are invariant under the rescaling of $\Lambda$.
Nevertheless,  the scheme-dependent grand  thermodynamic potential  transforms as
\bea\label{ffa}
\Omega(a,T,\mu)=a^4 f(\frac{a}{T},\frac{a}{\mu})+a^4\log(\frac{a}{\Lambda})\frac{N^2_c}{48 \pi^2}.
\eea
The entropy density $s$ evaluated at the horizon is scheme-independent
\bea
\frac{\partial s}{\partial \Lambda}=0.
\eea
 We stress that the $U(1)$ chemical potential $\mu$ is also scheme-independent. A consistent check can be made by using (\ref{chenmecial}) and the concrete expression  (\ref{mu1}). The rescaling
 \bea
 x_{i}=k x'_{i}, ~~v=k v',
 \eea  would not shift $\mu$ and there is no logarithmic term in the expression for $\mu$. The scheme-independence of $\mu$ is also implied by its thermodynamic definition (\ref{ca}) together with (\ref{ffa}). In contrast, the  potential $\Phi$ is scheme-dependent, which can be indicated both by the thermodynamic definition (\ref{phib}) and its $3$-form gauge potential definition.
Before we investigate the thermodynamic phase structure of this charged and anisotropic system, we shall write down the necessary and sufficient condition for local thermodynamic stability
\bea
&&c_{\rho,a}\equiv T\bigg(\frac{\partial s}{\partial T}\bigg)_{\rho,a}> 0,\\
&& \Phi'\equiv\bigg(\frac{\partial \Phi}{\partial a}\bigg)_{\rho,T}>0,\label{chemical}\\
&& \mu'\equiv\bigg(\frac{\partial \mu}{\partial \rho}\bigg)_{a,T}>0. \label{chemical1}
\eea
The heat capacity $c_{\rho,a}$ at constant charges $\rho$ and $a$ should be positive and regular. The second  and third conditions (\ref{chemical}) and (\ref{chemical1}) state  that the system is stable against infinitesimal ``charge" fluctuations.

\section{Phase structure}\label{phase}
The black hole solution found in asymptotically AdS space shows its rich phase structure. For neutral black holes with spherical topology in asymptotically AdS spacetime, there is a phase transition at a given temperature from
a description in terms of lower temperature AdS to a black hole setup \cite{hawking page}. This so called Hawking-Page transition is due to a competing effect between the scale set by the volume of the spacetime and the scale set by the temperature. When a $U(1)$ gauge field is included, the solution describes an asymptotically RN-AdS black hole with a horizon topology $S^{d-1}$, which gives rise to an interesting phase structure both in the canonical and grand-canonical ensembles \cite{van}.

We stress that the scale determined by the volume of the $S^{d-1}$ plays a key role in the phase diagrams. It turns out that the phase diagram of RN-AdS with $S^{d-1}$ horizon is analogous to the phase structure of van der Waals' liquid-gas system \cite{van,lu1,lu2,lu3,cai}. However, for the horizons with topology $\mathbb{R}^{d-1}$ which can be obtained by considering an infinite volume limit from the $S^{d-1}$ topology by using the transformation $r\rightarrow \eta r$ and $t\rightarrow \eta^{-1} t $ with $\eta \rightarrow \infty$, the black brane phase structures are usually considered to be trivial and the thermodynamics is dominated by the black holes for all temperatures. For example, the charged Lifshitz-like black brane solution was obtained numerically from the action \cite{comp5,comp6}
\bea
S'=\frac{1}{2\kappa^2}\int d^{d+2}x\sqrt{-g}\bigg(R+\frac{d(d+1)}{L^2}-2(\partial \phi)^2-e^{2 \alpha \phi}F^2\bigg).
\eea
The authors obtained the ``UV completion" by embedding black branes into asymptotic AdS space. In their papers, the Hawking temperature $T$ and chemical potential $\mu$ are two scales. It was proved that
in the Lifshitz-like regime $T\ll \mu$ and the AdS-like regime $T\gg \mu$, there were no thermal instabilities during the transition between the two regimes.

As to our case,  we deal with the anisotropic charged black brane whose horizon has $\mathbb{R}^{3}$ topology.  We will find that, unlike the isotropic  RN-AdS black brane with trivial phase structure, our anisotropic black brane solution has a new parameter of anisotropy which enriches the phase structure. On the other side, compared with the neutral anisotropic black brane solution, what is new in our solution is the presence of the $U(1)$ chemical potential. We will see later that the $U(1)$ chemical potential and charge density yielding no conformal anomaly will significantly modify the phase diagram set up in \cite{mateos}. So in the following, it is interesting to study the phase structure by comparing with the RN-AdS black brane and the chargeless anisotropic black brane, respectively. We will uncover that there are two kinds of thermodynamic instabilities for this  black brane system: One instability is scheme-independent and signals a Hawking-Page phase transition at a smaller horizon radius $r_c$; the other instability is scheme-dependent and is similar to that found in \cite{mateos}, which implies that the brane has the tendency to ``filamentation".

We   mainly explore the phase structure of anisotropic plasma with a finite chemical potential at finite temperature  in the small anisotropy limit. In the weak anisotropy limit,
the black brane solution is obtained analytically by perturbing around the isotropic RN-AdS solution.  The details can be found in Appendix \ref{highApp} and \ref{low App}. Note that the black brane temperature can be very low, although it cannot reach zero in our case unless $a$ takes imaginary values \footnote{For the case with imaginary anisotropic parameter $a$, see Appendix \ref{imag}.}. So the analytical solution given in Appendix \ref{low App} can cover both the low and high temperature regimes \footnote{The high temperature black brane solution and various quantities in the small charge density and weak anisotropy limits up to the $\mathcal{O}(a^4)$ order   are collected in  Appendix \ref{highApp}.}.

\subsection{Phase structure compared with isotropic RN-AdS black brane: scheme-independent instability}\label{sec41}
It is well-known that the isotropic RN-AdS black brane with planar horizon is thermodynamic stable  and the grand thermodynamic potential is negative for all temperatures.  However, when the anisotropy is presented, as shown in Fig.\ref{beta}, we find that there are qualitatively two distinct branches of solution for a given temperature because $\uh$ has two positive roots: a branch with larger radii and one with smaller.
 \begin{figure}[htbp]
 \begin{minipage}{1\hsize}
\begin{center}
%\vspace*{10mm}
\includegraphics*[scale=0.5] {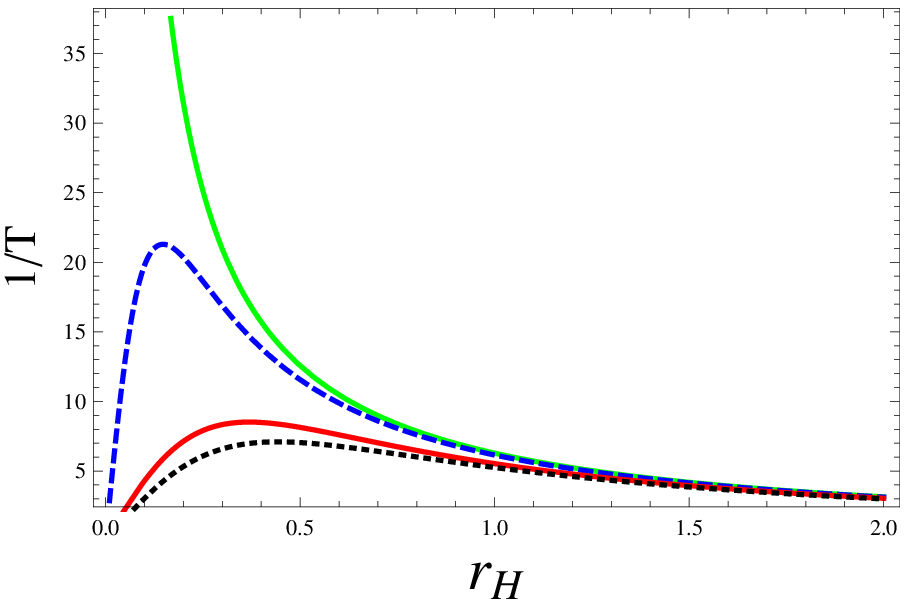}
\includegraphics*[scale=0.5] {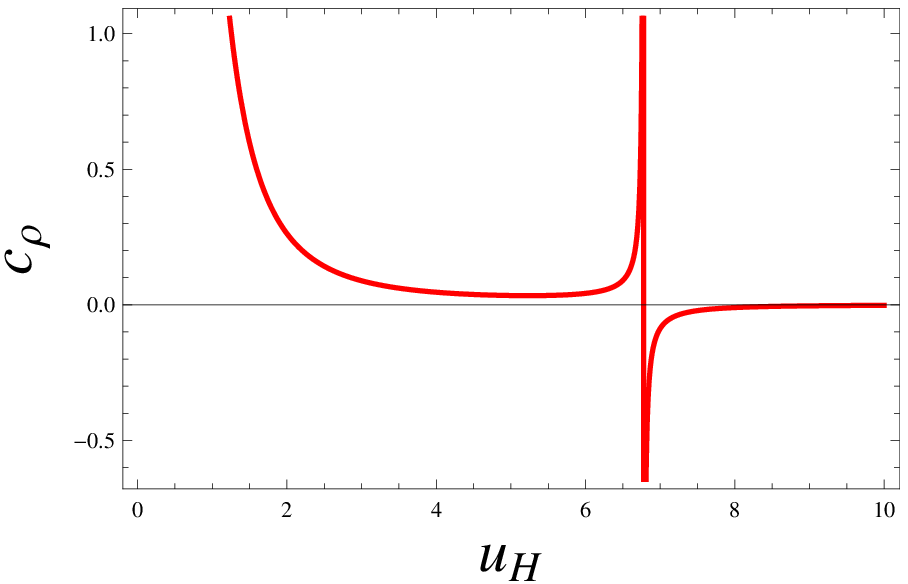}
\includegraphics*[scale=0.5] {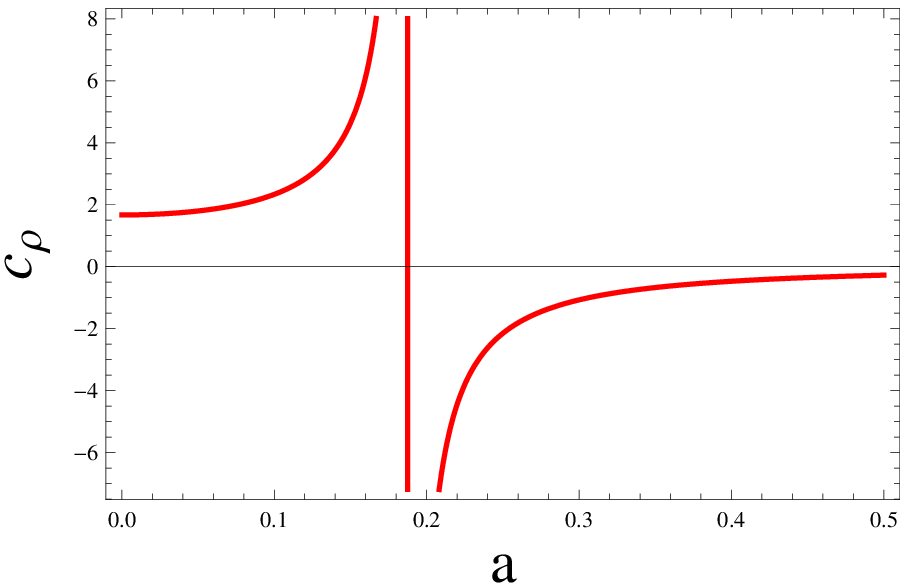}
\end{center}
\caption{(Color online.) On the left is a graph of the inverse temperature \textit{vs.} horizon radii for $q=1$ and $N_c=1$. The lines from top to bottom corresponds to $a=0$ (green), $a=0.2$(blue), $a=0.5$ (red) and $a=0.6$ (black), respectively. The center graph depicts the specific heat as a function of the inverse horizon radii $\uh$, where we set $N_c=1$, $a=0.2$, and $q^2=1$. On the right is the specific heat $c_{\rho}$ as a function of $a$, where we set $N_c=1$, $\uh=2$, and $q^2=1.8$. There is a sharp peak at $a=0.2$, signaling an instability.}\label{heat1}
\end{minipage}
\end{figure}
As a demonstration, we plot the inverse temperature of the black brane   as a function of the horizon radius $r_H=1/\uh$ with diverse $a$ in Fig.\ref{heat1} (left). It is easy to see that, except the top line    which corresponds to the  isotropic RN-AdS black brane, there exist two branches of black brane solutions associated with each temperature. The non-vanishing $a$ case is quite similar to the familiar case of the uncharged Schwarzschild-AdS black bole with $S^3$ horizon topology and spherical RN-AdS black hole at fixed charge.
This implies that the black brane  with $\mathbb{R}^{d-1}$ topology we presented  may have the non-trivial phase structure. Furthermore, we can see that the local slope of the $1/T$ curve is positive for the smaller radii branch, meaning that the temperature decreases as $r_{H}$ increases.
Therefore, the smaller branch with smaller radii is unstable, having negative specific heat (see Fig.\ref{heat1} (Middle and right) as an example). The smaller branch solution is unphysical and should not be applied to studying the dual CFT.

We conclude that the anisotropic charged black brane solution is unstable and there is a Hawking-Page phase transition at the horizon radii $r_c$. The instability uncovered here is due to the competing effect between the horizon radii and the anisotropy $a$\footnote{Note that both them have the dimensions of mass.}.  The instability can also be inferred from Fig.\ref{beta}  where  $\partial s/\partial T<0$ and the third law of thermodynamics is violated at this branch. We should emphasize the instability uncovered here is independent of the reference scale $\Lambda$ (  i.e. the renormalization scheme ), because both the temperature and the entropy density can be determined from the horizon values.

\subsection{Phase structure compared with neutral anisotropic black brane: scheme-dependent instability}
Now we are going to inspect what is different in the phase structure after adding the $U(1)$ chemical potential to the anisotropic black brane solution given in \cite{mateos}. We would like to assume that the following discussions are
carried out in the larger horizon radii branch (i.e. $c_{\rho,a}>0$ and $\mu'>0$), so that we can focus on the scheme-dependent thermodynamic variables.  Note that the energy density, pressure and the potential $\Phi$ are all scheme-dependent.

 As it was emphasized in \cite{mateos} that the presence of a conformal anomaly and the reference scale $\Lambda$ play a crucial role in the thermodynamics and the phase diagram. It was already noted that the coefficients of the $\mathcal{O}(a^4)$ terms, in particular the precise value of $c_{sch}$ and $c_{int}$, have no influence on the phase structure \cite{mateos}. In this sense, we will add the $\log(a/\Lambda)$ term into the stress tensor, although in  Appendix D, we only obtained the solution up to the  $\mathcal{O}(a^2)$ order. The reason is that, in the regions of $\Lambda\ll a$ and $\Lambda\gg a$, the term $\log(a/\Lambda)$  will play an important role in the sign of the energy density, the ``chemical" potential $\Phi$ and pressures. The discussions and the results below closely follow \cite{mateos} and the main difference is the presence of the $U(1)$ chemical potential, which will modify the phase structure.

 We  summarize the energy density and pressures and add the $\log(a/\Lambda)$ term  as follows
  \bea
 E&&=\frac{3(1+q^2)\nc^2}{8\pi^2\uh^4}+\frac{\nc^2\Big(-2\sqrt{1+4q^2}+5(q^2-2)\log\left(\frac{3-\sqrt{1+4q^2}}{3+\sqrt{1+4q^2}}\right)\Big)}{64\pi^2\sqrt{1+4q^2}\uh^2}a^2+\frac{N^2_c a^4}{48\pi^2}\log(\frac{a}{\Lambda}),\nonumber\\[0.3mm]
 P_{xy}&&=\frac{(1+q^2)\nc^2}{8\pi^2\uh^4}+\frac{\nc^2\Big(2\sqrt{1+4q^2}+5(q^2-2)\log\left(\frac{3-\sqrt{1+4q^2}}{3+\sqrt{1+4q^2}}\right)\Big)}{192\pi^2\sqrt{1+4q^2}\uh^2}a^2
 -\frac{N^2_c a^4}{48\pi^2}\log(\frac{a}{\Lambda}),\nonumber\\
 P_z&&=\frac{(1+q^2)\nc^2}{8\pi^2\uh^4}+\frac{5\nc^2\Big(-2\sqrt{1+4q^2}+(q^2-2)\log\left(\frac{3-\sqrt{1+4q^2}}{3+\sqrt{1+4q^2}}\right)\Big)}{192\pi^2\sqrt{1+4q^2}\uh^2}a^2+3\frac{N^2_c a^4}{48\pi^2}\log(\frac{a}{\Lambda}).\nonumber\\
  \eea
The ``chemical" potential with respect to the  ``charge" $a$ is given by
\bea
\Phi=-\frac{ \nc^2 a}{16 \pi^2 \uh^2}+4\frac{N^2_c a^3}{48\pi^2}\log(\frac{a}{\Lambda}).
\eea
However, the chemical potential related to the $U(1)$ field is independent of the reference scale $\Lambda$:
\bea\label{mu1}
\mu&&=\frac{q}{8\sqrt{3}\uh}\left(24+\frac{5\uh^2\log(\frac{3-\sqrt{1+4q^2}}{3+\sqrt{1+4q^2}})}{\sqrt{1+4q^2}}a^2\right).
\eea
This is also implied by (\ref{ca}). The absence of the $\log(a/\Lambda)$ term in the chemical potential will significantly change the phase structure of the system as we will see below. It is also clear that the chemical potential for this prolate black brane solution is less than the chemical potential of RN-AdS black brane for $a^2>0$.

In \cite{mateos}, the low-temperature regime corresponds to the Lifshitz-like geometry. In particular, the zero temperature $T=0$ requires $-g_{tt}=g_{xx}=g_{yy}$ and $\cf\cb=1$. For our case, there may also
exist a renormalization group flow between the AdS geometry in the ultraviolet and a Lifshitz-like geometry in the infrared. However, our black brane temperature can be low enough even  in the weak anisotropy limit in the ultraviolet. We will not rely on the Lifshitz-like  analysis here.

We are now considering the system consisting of different components (substances) and comprising two phases (i.e. isotropic phase and anisotropic phase).   Since the system contains different charges $a$ and $\rho$, we shall   take account of the mixture effects of the two substances. The coexistence of the two phases at equilibriums can be clarified by five conditions
\bea\label{coe}
\Omega-\Omega^0=0,~~~~\Phi=0,~~~P_z-P^0=0,~~~~\bigg(\frac{\partial\Phi}{\partial a}\bigg)_{T,\rho}=0,~~~\mu-\mu^0=0.
\eea
Before we focus on the effects of the $U(1)$ chemical potential to this anisotropic black brane phase diagram, let us first follow \cite{mateos} and discuss the physical meaning of the first four conditions in (\ref{coe}).
These four equations yield four solutions $a_i$, meaning that $\Omega,\Phi,P_z,\Phi'$ change sign at these values in the order in which they are listed above. These values satisfy
\bea
a_{F}>a_{\Phi}>a_{P_z}>a_{\Phi'}.
\eea
The exact values of $a_i$ depend on $N_c$, $\uh$ and $q$, but their ordering is independent of these constants. At low temperatures and low densities, $a<a_{\Phi'}$, the system is unstable against infinitesimal charge fluctuations. At densities $a_{\Phi'}< a<a_{P_z}$, the system stays in the metastable regime and it becomes stable against infinitesimal charge fluctuations, but is  still unstable against finite charge fluctuations. If $a<a_{P_z}$, the pressure of the isotropic phase is higher than that of anisotropic phase $P_z< P^0$. This could lead to the fact that bubbles of isotropic phase can form and grow inside the anisotropic phase. Therefore a homogeneous phase of density $a<a_{P_z}$ will fall apart into a mixed phase consisting of high-density anisotropic `droplets' or `filaments' of anisotropic phase with density $a=a_{P_z}$ and $P_z=0$ surrounded by vacuum regions with $a=P^0=0$ \cite{mateos}. This phenomena is very similar to what found in QCD at low $T$ and finite baryon density: The pressure of a chirally broken homogeneous phase with density lower than a critical density $n_0$ is negative and the role of the chirally restored phase is played by the anisotropic phase; the analogue of $n_0$ is $a_{P_z}$ and the`droplets'  correspond to the regions of non-zero D7-brane density \cite{QCD,QCD1,QCD2}. The similarities between the physics displayed in the region $a<a_{P_z}$ and that encountered in QCD at low temperature and finite baryon density was also elaborated in \cite{mateos}.

Now, we consider the fact that the presence of the $U(1)$ chemical potential and the anisotropy significantly change the phase diagram of the whole system. The chemical potential of the isotropic phase is higher than that of the
anisotropic phase $\mu^0>\mu$. Thus the fifth equation in (\ref{coe}) cannot be satisfied unless $a=0$.  This means that the anisotropic phase is more stable than the isotropic phase. As a consequence, charges or baryons would immigrate from the isotropic phase to the anisotropic phase, forcing a redistribution of the total charge and baryons. We would like to follow \cite{mateos} and divide the phase diagram into five distinct zones:
\begin{center}
~~~~~~~~~I: ~~$\Omega>\Omega^0$,~~$\Phi>0$,~~$P_z>P^0$,~~$\mu<\mu^0$,~~$\mu'>0$,~~$\Phi'>0$,\\
~~~~~~~~II: ~~$\Omega>\Omega^0$,~~$\Phi>0$,~~$P_z>P^0$,~~$\mu<\mu^0$,~~$\mu'>0$,~~$\Phi'>0$,\\
~~~~~~~III: ~~$\Omega<\Omega^0$,~~$\Phi<0$,~~$P_z>P^0$,~~$\mu<\mu^0$,~~$\mu'>0$,~~$\Phi'>0$,\\
~~~~~~~IV: ~~$\Omega<\Omega^0$,~~$\Phi<0$,~~$P_z<P^0$,~~$\mu<\mu^0$,~~$\mu'>0$,~~$\Phi'>0$,\\
~~~~~~~V: ~~$\Omega<\Omega^0$,~~$\Phi<0$,~~$P_z<P^0$,~~$\mu<\mu^0$,~~$\mu'>0$,~~$\Phi'<0$.\\
\end{center}
Different from the phase diagram given in \cite{mateos}, in the zones I, II and III, the thermodynamically preferred configuration is a metastable homogeneous phase  because the $U(1)$ chemical potential in the anisotropic regions  is not the same as that in isotropic regions. In zone $IV$, the homogeneous phase is metastable because $P_z<P^0$ and $\mu<\mu^0$. Zone V is identified as an unstable phase since it relates to the tendency of the charge $a$ to clump together and also the $U(1)$ charge $\rho$ has the tendency to escape from the isotropic phase to the anisotropic phase.

The $U(1)$ chemical potential at phase equilibrium  should be the same in the isotropic and anisotropic regions in that a charged black brane can be considered as a system with an infinite charge reservoir and the chemical potential eventually equilibrates to the same value everywhere.  In contrast, the ``chemical potential" $\Phi$ need not to be the same in these phases: $\Phi$ vanishes in the isotropic phase but non-zero in the anisotropic phase.
This is an important consistency condition for the coexistence of the two phases: $\Phi_{ani}\leq \Phi_{iso}$, since otherwise the D7-brane charge would escape from the anisotropic to the isotropic regions.

\section{Summary and Outlook}
\label{discussion}
In summary, in the spirit of the applications of gauge/string duality in QCD, we have studied the five dimensional anisotropic black brane solution in Einstein-Maxwell-dilaton-axion theory from the type IIB supergravity theory.  The solution we presented obeying $AdS_5\times S^5$ boundary conditions and possessing a regular anisotropic horizon, is a RN-AdS version of the anisotropic black brane solution obtained in \cite{mateos}. What is new in this paper is that we introduce the $U(1)$ gauge field, which corresponds to conserved number operators in the dual field theory. The numerical and analytical computations show that the extremal black brane limit can be reached only if the anisotropic parameter $a$ becomes zero or imaginary. The entropy
density does not vanish as the temperature becomes zero.

We also studied the thermodynamics of this anisotropic system for the case $a^2>0$. Several black brane thermodynamic quantities were obtained and the first law of black hole thermodynamics holds.
The stress  tensor and the holographic renormalization are calculated in the presence of the $U(1)$ gauge field. While the energy density, the potential $\Phi$ and pressure transforming under the rescaling
$(a,T)\rightarrow (ka,kT)$ contain  an inhomogeneous piece caused by the presence of a non-zero conformal anomaly $\mathcal{A}=N^2_c a^4/48\pi^2$. The chemical potential $\mu$ is scheme-independent and yields no logarithmic term under the scale rescaling $x_{i}\rightarrow k x_{i} $ and $v\rightarrow k v$. This fact indicates that the physics not only depends on the scale $T$, $\mu$ and $a$, but also depends on an additional scale $\Lambda$.

We analyzed the phase structure of this ``prolate" system  by comparing with those of isotropic RN-AdS black brane and uncharged anisotropic black brane, respectively. Intriguingly, we uncovered scheme-independent instability by exploring the temperature-horizon radii relation and the behavior of the entropy. At a fixed temperature, there are  two distinct branches of black brane solution: a branch with larger radii and one with smaller. The smaller branch is unstable, yielding a negative specific heat. The phase structure of this ``prolate"  black brane solution is similar to Schwarzschild-AdS black hole with spherical horizon. The instability found here
signals a competing effect between the horizon radius and the anisotropy constant.

Even when we turned to the larger radii branch and investigated the scheme-dependent thermodynamic variables, we found that there are only metastable and unstable zones of the phase diagram because of $\mu< \mu^0$. As discussed in section \ref{phase}, zones $\rm I, II, III$ and $IV$ are metastable. That is to say, the system is unstable against finite $U(1)$ charge fluctuations. Zone $V$ is   metastable against finite $U(1)$ charge fluctuations, but unstable to infinitesimal  charge $a$ fluctuations.  It is worth future investigations on the similarities between  the anisotropic phase diagram obtained here and the anisotropic QGP behavior discovered in the experiments.

In the future work,   we would like to report  on the effects of the anisotropy and the $U(1)$ chemical potential on several observables. The dual anisotropic holographic fermions will be investigated.  The jet quenching parameter, the drag force and heavy quark energy loss \cite{jet1,jet2,drag1,drag2} due to the anisotropic effects at finite $U(1)$ chemical potential will also be studied. The study on transport coefficients of the dual anisotropic hydrodynamics, in particular the conductivity and shear viscosity  are also in progress.

%%%%%%%%%%%%%%%%%%%%%%%%%%%
\section*{Acknowledgements}

We are indebted to Hong Lu and  Jian-Xin Lu for help on the Type IIB supergravity action. We would like also to thank  Jia-Rui Sun,  Shang-Yu Wu, Yi Yang,  Li Qing Fang, Xiao-Mei Kuang and Shao-Jun Zhang   for helpful discussions. We also acknowledge Ying Jiang and Guo-Hong Yang for indispensable encouragements. XHG was partially supported by NSFC,
China (No.11375110),  and Shanghai Rising-Star Program (No.10QA1402300). SJS was partially supported by the NRF grant funded by the
Korea government(MEST) through the Mid-career Researcher
Program with grant NRF-2013R1A2A2A05004846.

%%%%%%%%%%%%%%%%%%%%%%%%%%
%\newpage
\appendix

%%%%%%%%%%%%%%%%%%%%%%%%%%

\section{Derivation of the solution}
\label{derivation}
\subsection{The equations of motion}
In this appendix, we   give the detailed process of derivations of our solution.
We start with five-dimensional equations of motion in explicit form:
the dilaton equation
 \bea
\frac{1}{\sqrt{-g}}\partial_\mu\left(\sqrt{-g}\partial^\mu\phi\right)-e^{2\phi}(\partial\chi)^2=0\,,
\label{dilatonEOM1}
\eea
the Maxwell equations
\begin{eqnarray}
\frac{1}{\sqrt{-g}}\partial_\mu\left(\sqrt{-g}\partial^\mu A^\nu\right)=0 ,
\label{MaxwellEOM1}
\end{eqnarray}
and the Einstein equations
\begin{eqnarray}
R_{\mu\nu}-\frac{1}{2}\partial_{\mu}\phi\partial_{\nu}\phi-\frac{e^{2\phi}}{2}\partial_{\mu}\chi\partial_{\nu}\chi-\frac{1}{2}F_{\mu\lambda}F_\nu^{~\lambda}+\frac{g_{\mu \nu}}{12}F_{\lambda\rho}F^{\lambda\rho}+4g_{\mu\nu}=0~~.
 \label{EinsteinEOM1}
\end{eqnarray}

We suppose the metric is of the asymptotic $AdS_5$ form
\bea
ds_5^2 =  \frac{e^{-\frac{1}{2}\phi}}{u^2}\left( -\cf \cb\, dt^2+dx^2+dy^2+ \ch dz^2 +\frac{ du^2}{\cf}\right).
 \label{10dmetric}
\eea
One can simplify the following calculations further by setting
\bea
\ch=e^{-\phi}\,.
\label{simplifying_ansatz}
\eea
The axion field is taken as
\bea
\chi(z)= a \, z\,.
\label{chi}
\eea
Then, the explicit expressions for the equations of motion can be obtained straightforwardly after plunging the ansatz into (\ref{dilatonEOM1})-(\ref{EinsteinEOM1}).
At first, we can express the Maxwell equation as
\bea
\partial_u\Big(\frac{e^{-\frac{7}{4}\phi}\sqrt{\cb}}{u^5}F^{ut}\Big)=0 ,
\eea
 which results in the $U(1)$ gauge field as
\bea
A_{t}(u)=-\int^u_{\uh}Q\sqrt{\cb}e^{\frac{3}{4}\phi}u du ,
\label{potential}
\eea
where the $Q$ is an integral constant, and the requirement of the vanishing gauge field at horizon $A_t(\uh)=0$ is imposed.
Then, the dilaton and Einstein equations read
\bea
0&=&-e^{7\phi/2}a^2-\frac{3e^{\phi/2}\cf\phi'}{u}+\frac{e^{\phi/2}\cf\cb'\phi'}{2\cb}+e^{\phi/2}\cf'\phi'-\frac{5e^{\phi/2}\cf\phi'^2}{4}+e^{\phi/2}\cf\phi'',
\label{eq4}\\
\cr
0&=&-\frac{8e^{-\phi/2}}{u^2}-\frac{2e^{2\phi}u^4}{3}Q^2+\frac{8\cf}{u^2}-\frac{4\cf\cb'}{u\cb}-\frac{\cf\cb'^2}{2\cb^2}-\frac{5\cf'}{u}+\frac{3\cb'\cf'}{2\cb}\nonumber\\
&&~~~~~~~~+\frac{4\cf\phi'}{u}-\frac{3\cf\cb'\phi'}{2\cb}-\frac{7\cf'\phi'}{4}+\frac{5\cf\phi'^2}{8}+\frac{\cf\cb''}{\cb}+\cf''-\frac{\cf\phi''}{2}, \label{eq0}
\\
&&\cr
0&=&-\frac{4}{u}+\frac{4e^{-\phi/2}}{u\cf}-\frac{e^{2\phi}u^5}{6\cf}Q^2+\frac{\cb'}{2\cb}+\frac{\cf'}{\cf}-2\phi'+\frac{u\cb'\phi'}{8\cb}+\frac{u\cf'\phi'}{4\cf}-\frac{5u\phi'^2}{16}+\frac{u\phi''}{4},\label{eq1}
\\
&&\cr  0&=&-\frac{8}{u}+\frac{8e^{-\phi/2}}{u\cf}-\frac{e^{3\phi}u}{\cf}a^2-\frac{e^{2\phi }u^5}{3\cf}Q^2+\frac{\cb'}{\cb}\nonumber\\
&&~~~~~~~+\frac{2\cf'}{\cf}-7\phi'+\frac{3u\cb'\phi'}{4\cb}+\frac{3u\cf'\phi'}{2\cf}-\frac{15u\phi'^2}{8}+\frac{3u\phi''}{2},
\label{eq2}
\\
&&\cr
0&=&-\frac{16}{u^2}+\frac{16e^{-\phi/2}}{u^2\cf}+\frac{4e^{2\phi}u^4}{3\cf}Q^2+\frac{2\cb'}{u\cb}+\frac{\cb'^2}{\cb^2}+\frac{10\cf'}{u\cf}\nonumber\\
&&~~~~~~~ -\frac{3\cb'\cf'}{\cb\cf}-\frac{2\phi'}{u}+\frac{\cb'\phi'}{2\cb}+\frac{7\cf'\phi'}{2\cf}-\frac{7\phi'^2}{2}-\frac{2\cb''}{\cb}-\frac{2\cf''}{\cf}+6\phi''
\label{eq3}
\,,
\eea
where the  prime $'$ denotes derivatives with respect to $u$. Note that there are four components of Einstein equations since the $x-$ and $y-$ directions obey the same one.  Using (\ref{eq0}) and (\ref{eq3}), one can reduce (\ref{eq4}), (\ref{eq1}) and (\ref{eq2}) to three first-order equations:
\bea 0=&&-e^{7\phi/2}a^2+\frac{6e^{\phi/2}\cf\cb'}{5u\cb}-\frac{21e^{\phi/2}\cf\phi'}{5u}+\frac{e^{\phi/2}\cf\cb'\phi'}{\cb}+e^{\phi/2}\cf'\phi'-\frac{4e^{\phi/2}\cf\phi'^2}{5}
\label{con3},
\\
&&\cr
0=&&-\frac{8}{u}+\frac{8e^{-\phi/2}}{u\cf}-\frac{e^{3\phi}u}{\cf}a^2-\frac{e^{2\phi}u^5}{3\cf}Q^2\nonumber\\
&&+\frac{14\cb'}{5\cb}+\frac{2\cf'}{\cf}-\frac{44\phi'}{5}+\frac{3u\cb'\phi'}{2\cb}+\frac{3u\cf'\phi'}{2\cf}-\frac{6u\phi'^2}{5}
\label{con2},
\\
&&\cr
0=&&-\frac{4}{u}+\frac{4e^{-\phi/2}}{u\cf}-\frac{e^{2\phi} u^5}{6\cf}Q^2+\frac{4\cb'}{5\cb}+\frac{\cf'}{\cf}-\frac{23\phi'}{10}+\frac{u\cb'\phi'}{4\cb}+\frac{u\cf'\phi'}{4\cf}-\frac{u\phi'^2}{5}
\label{con1},
\eea
which leads to three equations of motion for $\cf(u), \cb(u), \phi(u)$ respectively:
\bea
\cf&=&\frac{e^{-\frac{1}{2}\phi}}{12(\phi'+u\phi'')}\left(3a^2 e^{\frac{7}{2}\phi}(4u+u^2\phi')+48\phi'-2e^{\frac{5}{2}\phi}Q^2u^6\phi'\right) \label{eq_F_app},\\
\frac{\cb'}{\cb}&=& \frac{1}{24+10 u\phi'}\left(24\phi'-9u\phi'^2+20u\phi''\right),\label{eq_B_app} \\
 0&=&\frac{-48 \phi '^2 \left(32+7 u \phi '\right)+768 \phi ''+4 e^{\frac{5 \phi }{2}} Q^2 u^5 \left(-24 \phi '+u^2 \phi
 '^3-8 u \phi ''\right)}{48 \phi '-2 e^{\frac{5 \phi}{2}} Q^2 u^6 \phi '+3 a^2 e^{\frac{7 \phi}{2}} u \left(4+u \phi '\right)}
 +\frac{1}{u \left(12+5 u \phi '\right) \left(\phi '+u \phi ''\right)}\nonumber\\
 && \times\Big[13u^3\phi'^4+u^2\phi'^3(96+13u^2\phi'')+8u(-60\phi''+11u^2\phi''^2-12u\phi^{(3)}) \nonumber\\
 &&~~~~~+2u\phi'^2(36+53u^2\phi''-5u^3\phi^{(3)})+\phi'(30u^4\phi''^2-64u^3\phi^{(3)}-288+32u^2\phi'')\Big].
 \label{3order_dil}
\eea
Considered a special case of isotropy (i.e. $a=0$), it is easy to observe that the dilaton equation with the boundary conditions of
$\phi(0)=0, \phi'(\uh)=0, \phi''(\uh)=0$ has a trivial solution $\phi=0$ . This fact will be important for the following analytic discussions.

However, in the presence of the axion filed, these equations of motion can not be solved easily by analytic method except for some special cases which will be considered in appendix \ref{highApp}. Now we would like to carry out the numeric analysis first.

%%%%%%%%%%%%%%%%%%

\subsection{Numeric analysis: ``prolate" solution}
\label{numerical}
To solve third-order equation numerically, like procedure used in \cite{mateos}, it is more convenient to shift the dilaton
\bea
\phi \to \tilde\phi\equiv \phi+\frac{4}{7}\log a\,,
\label{dilaton_shifted}
\eea
and eliminate $a$ from (\ref{3order_dil}) altogether. Note that the anisotropy constant $a$ is assumed to be a real and positive number here. The associated boundary conditions are imposed as follows:  Inserting $\cf(\uh)=0$ in (\ref{eq_F_app}) and (\ref{3order_dil}),  one can obtain
\bea\label{bcondition}
\tilde\phi'(\uh) &=& -\frac{12a^{10/7}e^{7\tilde\phi_H/2}\uh}{-2e^{5\tilde\phi_H/2}Q^2\uh^6+3a^{10/7}\left(16+e^{7\tilde\phi_H/2}\uh^2\right)} \,,\nonumber\\
&&\cr
\tilde{\phi}''(\uh) &=&\frac{36a^{10/7}e^{6\tilde\phi_H}\uh^2\left(4e^{5\tilde\phi_H/2}Q^4\uh^{10}-12a^{10/7}Q^2\uh^4\left(8+e^{7\tilde\phi_H/2}\uh^2\right)+3a^{20/7}e^{\tilde\phi_H}\left(128+e^{7\tilde\phi_H/2}\uh^2\right)\right)}{\Big(-2e^{5\tilde\phi_H/2}Q^2\uh^6+3a^{10/7}\big(16+e^{7\tilde\phi_H/2}\uh^2\big)\Big)^3}
\nonumber\\
&&\cr
\tilde{\phi}'''(\uh) &=&\frac{72a^{10/7}e^{7\tilde\phi_H/2}}{\uh \left(2 e^{\frac{5 \tilde{\phi }_H}{2}} Q^2 \uh^6-3 a^{10/7} \left(16+e^{7\tilde\phi_H/2}\uh^2\right)\right)^5}\nonumber\\
&&\times\Big[32 e^{10 \tilde{\phi}_H} Q^8 \uh^{24}-64 a^{10/7} e^{15\tilde{\phi}_H/2} Q^6 \uh^{18} \left(17+3 e^{7\tilde{\phi}_H/2}\uh^2\right)+4a^{20/7}e^{5\tilde\phi_H}Q^4\uh^{12}(-1152\nonumber\\
&&+3376e^{7\tilde\phi_H/2}\uh^2+75e^{7\tilde\phi_H}\uh^4)-12a^{30/7}e^{5\tilde\phi_H/2}Q^2\uh^6\big(-27648+18880e^{7\tilde\phi_H/2}\uh^2+2036e^{\tilde\phi_H}\uh^4\nonumber\\
&&+3e^{21\tilde\phi_H/2}\uh^6\big)+9a^{40/7}\left(-98304+34816e^{7\tilde\phi_H/2}\uh^2+52864e^{7\tilde\phi_H}\uh^4-336e^{21\tilde\phi_H/2}\uh^6+3e^{14\tilde\phi_H}\uh^8\right)\Big]\nonumber
\\
\eea
After integrating (\ref{3order_dil}) numerically we use $\tilde\phi$ to  obtain $\cf$ and $\cb$ through (\ref{eq_F_app})-(\ref{eq_B_app}).

It is clear that the solution is determined by three parameters, $\tilde\phi_\textrm{H}$, $Q$ and $\uh$. This is what we expected, since these determine the three physical parameters that the solution must depend on  the temperature, the $U(1)$ charge density and the anisotropy.

By using the temperature formula (\ref{temperature}), we obtain
\bea
T =\sqrt{\cb_H}\left(-\frac{e^{2\tilde\phi_H}\uh^5 Q^2}{24\pi a^{8/7}}+a^{2/7}\frac{e^{-\tilde\phi_H/2}}{16\pi \uh}\left(16+e^{7\tilde\phi_H/2}\uh^2\right)\right),
\eea
where we have used the expression for $\cf'_H$:
\bea
\cf'_H =\frac{e^{2\tilde\phi_H}\uh^5 Q^2}{6a^{8/7}}-a^{\frac{2}{7}}\left(\frac{4e^{-\tilde\phi_H/2}}{\uh}+\frac{e^{3\tilde\phi_H}\uh}{4}\right).
\label{f1}
\eea
The above temperature can reduce to the RN-AdS black brane temperature in the $\phi \rightarrow0$ and $a \rightarrow 0$ limit,
 \bea
 T=\frac{1}{2\pi \uh}(2-q^2),
 \eea
 where $q=\frac{u^3_H Q}{2\sqrt{3}}$.
  For sake of simplicity of the computation, we would like to introduce a parameter $\overline{Q}$ defined as
\bea
\overline{Q}= a^{\frac{-5}{7}}Q.
\eea
The Hawking temperature becomes
\bea
T =a^{2/7}\sqrt{\cb_H}\left(-\frac{e^{2\tilde\phi_H}\uh^5 \overline{Q}^2}{24\pi}+\frac{e^{-\tilde\phi_H/2}}{16\pi \uh}\left(16+e^{7\tilde\phi_H/2}\uh^2\right)\right).
\eea
In principle, the zero temperature can be reached for nonzero $\sqrt{\cb_H}$, if the maximal value of $\overline{Q}$ is given by
\bea
\overline{Q}_{ext}=\sqrt{\frac{3}{2\uh^{6}}}\sqrt{16e^{-\frac{5}{2}\tilde\phi_H}+e^{\tilde\phi_H}\uh^2},
\eea
which reflects that larger $\tilde\phi_H$ leads to smaller $\overline{Q}_{ext}$. In absence of the dilaton, the maximal value of the charge is $Q=2\sqrt{6}\uh^{-3}$ corresponding
 to $q=2$.

\begin{figure}[htbp]
 \begin{minipage}{1\hsize}
\begin{center}
%\vspace*{10mm}
\includegraphics*[scale=0.50] {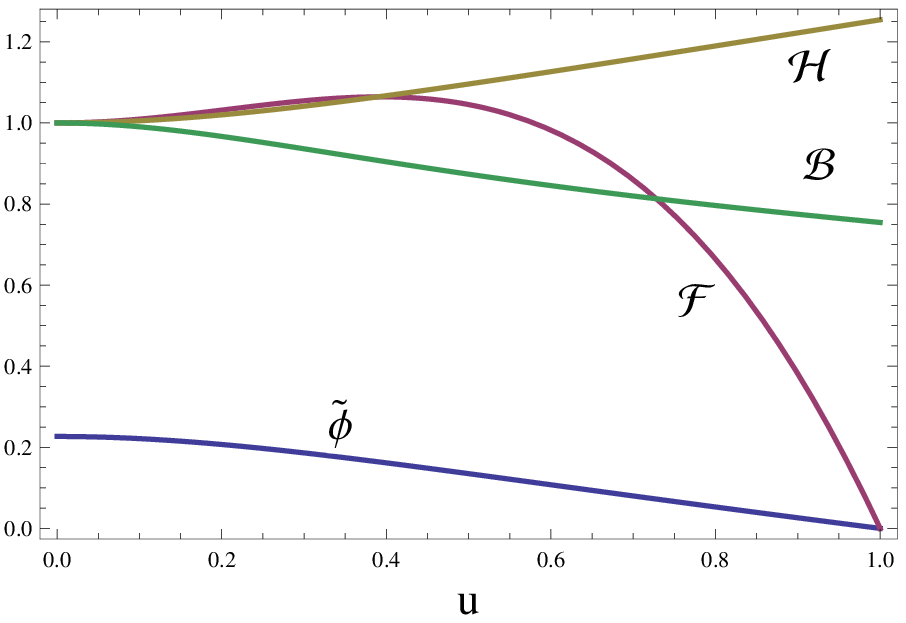}
\includegraphics*[scale=0.50] {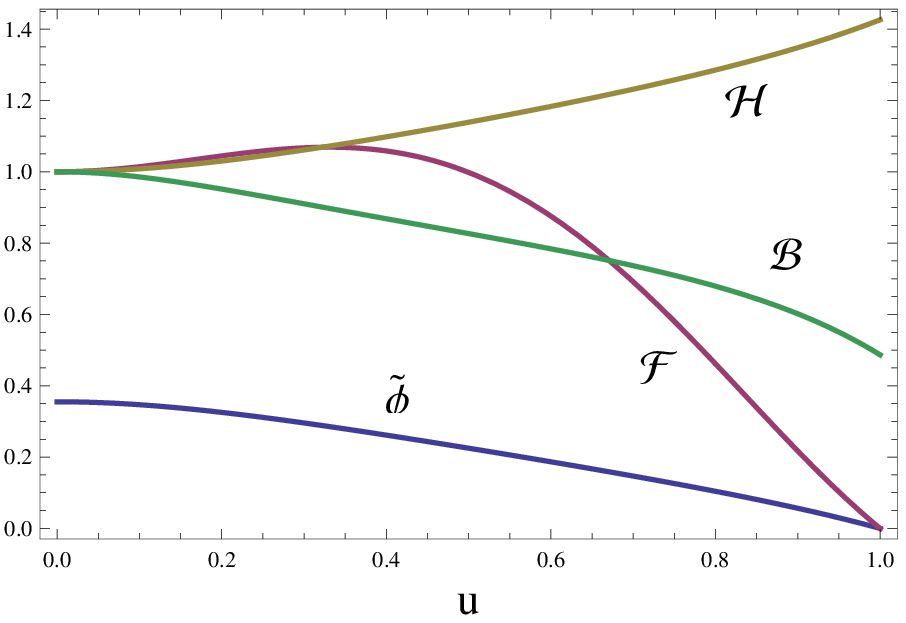}
\includegraphics*[scale=0.50]{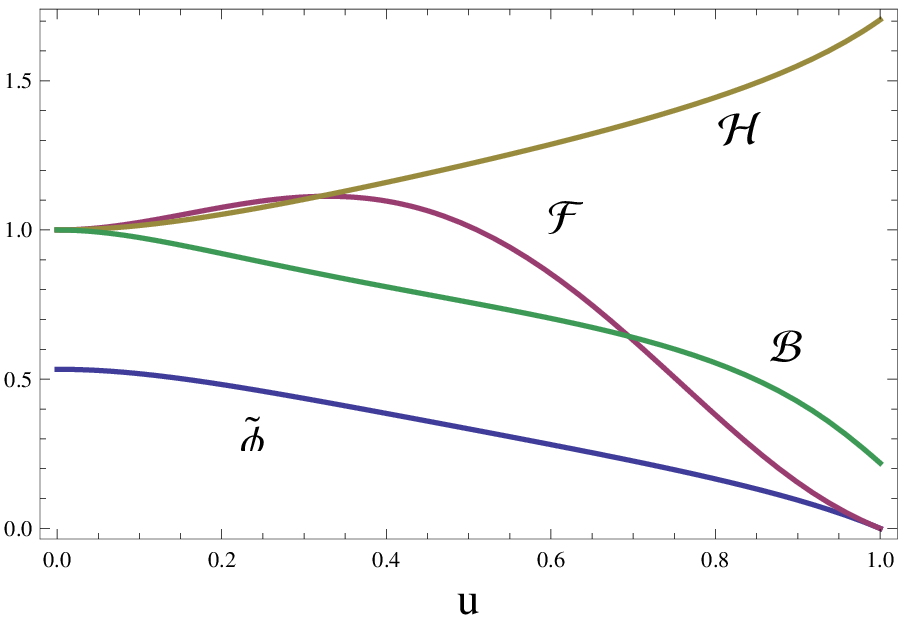}
\end{center}
\caption{(color online)The metric function for $\overline{Q}=1.5$ (left),  $\overline{Q}=4$ (middle) and $\overline{Q}=4.5$ (right), with $\tilde\phi_H=0$ and $\uh=1$.} \label{sol1}
\end{minipage}
\end{figure}

\begin{figure}[htbp]
 \begin{minipage}{1\hsize}
\begin{center}
%\vspace*{10mm}
\includegraphics*[scale=0.50] {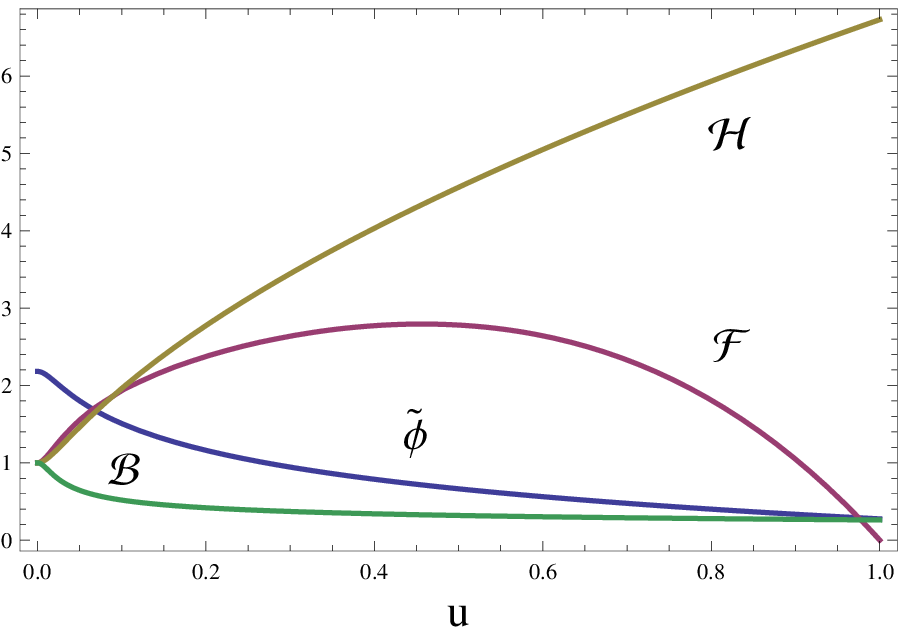}
\includegraphics*[scale=0.50]{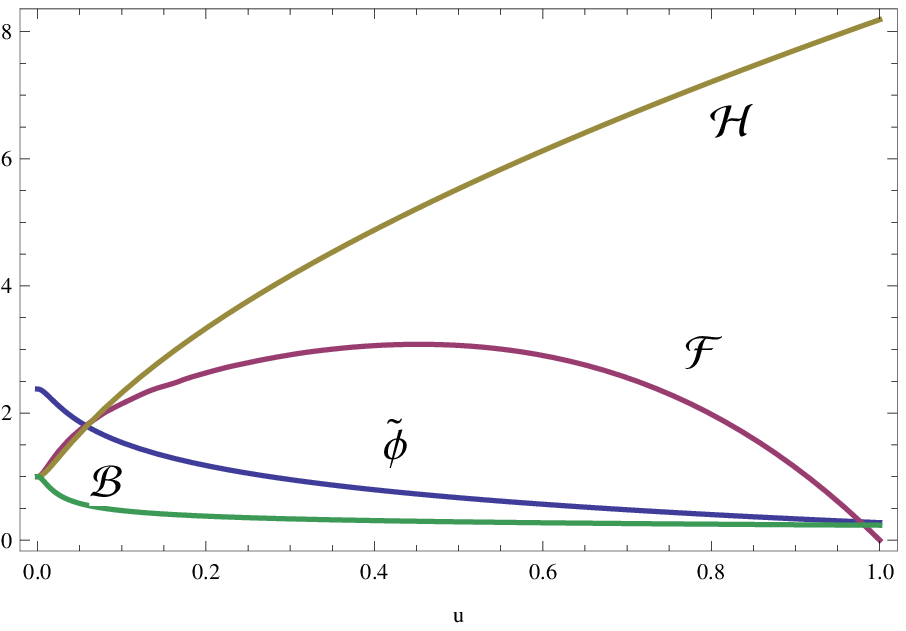}
\includegraphics*[scale=0.52] {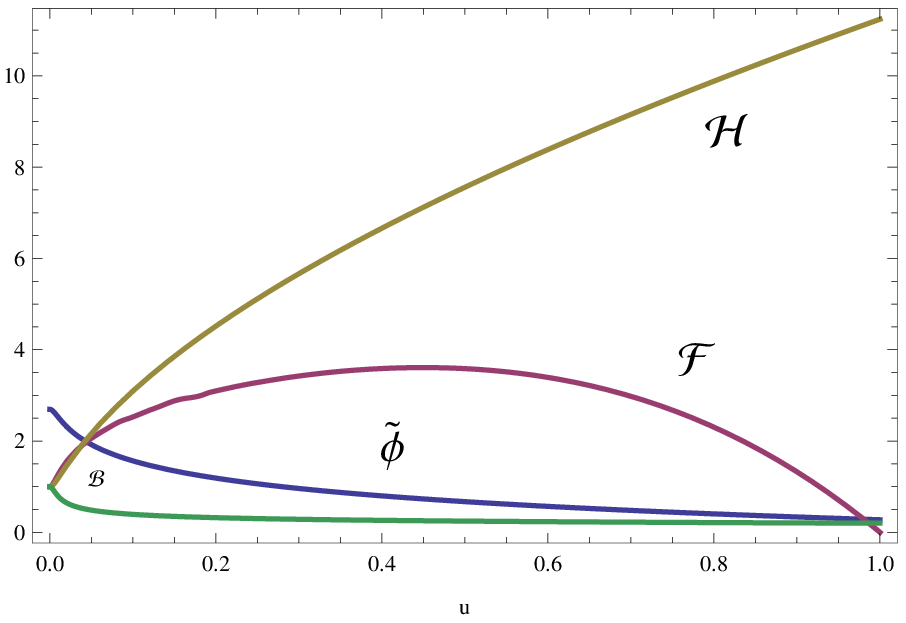}
\end{center}
\caption{(color online)The metric function for $\overline{Q}=0.25$ (left),  $\overline{Q}=0.5$ (middle) and $\overline{Q}=0.65$ (right), with $\tilde\phi_H=0.275$ and $\uh=1$.} \label{sol11}
\end{minipage}
\end{figure}

We explore the relations among $T$, $\overline{Q}$ and $a$ for different initial values of $\tilde\phi_H$ and $u_H$. We find that $T$ is sensitive to $\overline{Q}$  while  $a$ depends strongly on $\tilde\phi_H$. The profile for $\cb$ is seriously suppressed at the horizon for larger $\overline{Q}$ and thus the temperature becomes lower   because   the temperature is proportional to $\sqrt{\cb_H}$ (i.e. $T\propto \sqrt{\cb_H}$). In order to compare the result with the work of Mateos and Trancanelli \cite{mateos}, we solve the black brane solution numerically and present the plots in Fig.\ref{sol1} and Fig.\ref{sol11}. We mainly plot six different initial conditions: ($\tilde\phi_H=0, \uh=1, \overline{Q}=1.5, 4, 4.5$), which yield  an anisotropy-to-temperature ratio $a/T=4.96,17.73,59.75$, and ($\tilde\phi_H=0.275, \uh=1, \overline{Q}=0.25,0.5,0.65$), which yield  an anisotropy-to-temperature ratio $a/T=93.12,127.23,208.03$. We emphasize that in all the cases, the horizon value of $\ch(u)$ is greater than 1.
That is why this solution is called a ``prolate" solution.

\section{Holographic renormalization}
\label{Hrenormalization}
In this appendix, we   give a brief procedure for calculations of holographic renormalization for charged axion-dilaton-gravity system. Instead of Hamiltonian approach, we   carry out the traditional method \cite{anomaly}. The former formalism for chargeless case can be found in \cite{Yiannis}.

As in \cite{anomaly}, the metric near the conformal boundary can be expanded in Fefferman-Graham form
\bea \label{coord}
ds^2=g_{\mu\nu} dx^{\mu} dx^{\nu} =\frac{1}{r^2}\left(dr^2+h_{ij}(r)dx^i dx^j\right).
\eea
The Einstein equations in this coordinate  are given by
\bea
0&=&-\frac{1}{2}\mbox{Tr}(h^{-1}h'')+\frac{1}{4}\mbox{Tr}(h^{-1}h'h^{-1}g')+\frac{1}{2r}\mbox{Tr}(h^{-1}h')+\frac{4}{r^2}(h_{rr}-1)-\frac{1}{2}\phi'^2-\frac{r^2}{3}h_{tt}^{-1}A'^2 , \nonumber\\
0&=&R_{ij}(h)-\frac{1}{2}h_{ij}''+\frac{1}{2}(h'h^{-1}h')_{ij}+\frac{1}{2r}h_{ij}\mbox{Tr}(h^{-1}h')+\frac{3h'_{ij}}{2r}-\frac{1}{4}h_{ij}'\mbox{Tr}(h^{-1}h')\nonumber\\
&-&\frac{e^{2\phi}}{2}\partial_i\chi\partial_j\chi-\frac{r^2}{2}\partial_r A_i\partial_r A_j+\frac{r^2}{6}A'^2 h_{tt}^{-1}h_{ij},   \label{FGeinstein}
\eea
where the prime denotes differentiation with respect to $r$,  and $R_{ij}(h)$ is the Ricci tensor of $h_{ij}$.
The dilaton equation can be recast as
\bea
-3r\phi'+\frac{r^2}{2}(\log h)'\phi'+r^2\phi''-e^{2\phi}(\partial\chi)^2=0.
\eea
The Maxwell equation is given by
\bea
\left(\frac{\sqrt{h}}{r}h^{-1}_{tt} A_t'\right)'=0 ,\label{FG Max}
\eea
where $h$ denotes the absolute value of determinant of $h_{ij}$ for simplicity.

One may expand the metric as
\bea
h_{ij}(r)=h_{(0)ij}+r^2 h_{(2)ij}+r^4 h_{(4)ij}+2r^4 \tilde{h}_{(4)ij}\log r+\cdots,
\eea
the dilaton field as
\bea
\phi(r)=\phi_{(0)}+r^2\phi_{(2)}+r^4(\phi_{(4)}+2\psi_{(4)}\log r)+\cdots,  \label{diaton exp}
\eea
the gauge fields as
\bea
A_\mu(r)=A_{(0)\mu}+r^2A_{(2)\mu}+r^4( A_{(4)\mu}+2\tilde{A}_{(4)\mu}\log r)+\cdots,  \label{Max exp}
\eea
and the axion field as
\bea
\chi(r,z)=\chi_{(0)} + r^2 \chi_{(2)}+r^4 \left( \chi_{(4)} +2\tilde{\chi}_{(4)}\log r \ \right) +\cdots.  \label{axion exp}
\eea
Note that we can simply set $h_{(0)ij}=\eta_{ij}$ in our case. Then E.O.M (\ref{FGeinstein})-(\ref{FG Max}) can be solved order by order in $r$ and the results are shown in section \ref{sec-thermodynamics}.

 To obtain the  regulated action, we restrict the bulk integral to the region $\epsilon>0$ and then evaluate the on-shell action for the solution above.  We obtain
\bea
S_{reg}[\epsilon]&=&\frac{1}{2\kappa^2}\int d^4x\left[\int_\epsilon dr \sqrt{h} (-8)+\sqrt{\gamma} (\frac{1}{3}F^{\mu\nu}A_\mu n_\nu+2K)\right]\nonumber\\
&=&\frac{1}{2\kappa^2}\int d^4x\sqrt{h_{(0)}}\left(\epsilon^{-4}a_{(0)}+\epsilon^{-2}a_{(2)}+a_{(4)}\log\epsilon\right)+S_{fin},
\eea
where the first term in integral form above is divergent as $\epsilon\rightarrow0$. One can cancel the divergences by adding the counterterm that is defined as divergent part of  $-S_{reg}[\epsilon]$ in terms of fields on cut-off surface $r=\epsilon$.  The term $\gamma_{ij}=h_{ij}/\epsilon^2$ is the induced metric at cut-off surface whose extrinsic curvature is denoted as $K$ and   $\gamma$ is the absolute value of the determinant of $\gamma_{ij}$.

After some calculations, we obtain the coefficients of divergent terms
\bea
a_{(0)}=6 ,~~a_{(2)}=0,~~ a_{(4)}=-\frac{13a^4}{144}+4\mbox{Tr}h_{(4)ij}=-\frac{a^4}{6} ,
\eea
 where we have used $\sqrt{h_{(0)}}=1$. The coefficient $a_{(4)}$ is equal to  the conformal anomaly of the dual CFT \cite{anomaly}, which can be understood by noting that  the variation of the finite term must have the form
\bea
\delta S_{fin}=-\int d^4x \sqrt{h_{(0)}}\cal A\delta\sigma,
\eea
with $\mathcal{A}=-a_{(4)}/2\kappa ^{2}=\frac{a^4}{12\kappa^2}$   under the scale transformations  $\delta h_{(0)ij}=2\delta\sigma h_{(0)ij}$, $\delta\epsilon=2\delta\sigma\epsilon$ to guarantee the scale invariance.
Observe that all the divergent terms are exactly the same with chargeless case, therefore no additional counterterms are required to cancel the divergences in minimal subtraction. So the counterterm   is given by
\begin{eqnarray}
S_{ct}=-\frac{1}{\kappa^2} \int d^4x \sqrt{\g}
\left( 3- \frac{1}{8} e^{2\phi}\partial_i\chi\partial^i\chi \right)
+ \log r  \int d^4x \sqrt{\g} {\cal A}\,,
\end{eqnarray}
where ${\cal A}(\gamma_{ij},\phi,\chi,A_t)$ is the conformal anomaly in the Maxwell-axion-dilaton-gravity  system which satisfies $\lim_{r\rightarrow 0} \sqrt{\g} \, {\cal A}(\gamma_{ij},\phi,\chi,A_t)=\frac{a^4}{12\kappa^2}$.  Clearly, one can add any finite counterterms in the action, for instance $\frac{1}{4}(c_{sch}-1) \int d^4x \sqrt{\g} {\cal A}$ which corresponds to the renormalization scheme. Finally, recall that
the renormalized action is defined by
\begin{eqnarray}
S_{ren}=\lim_{\epsilon\rightarrow0} (S_{reg}+S_{ct}),
\end{eqnarray}
then the expect value of stress tensor is given by \cite{kostasgravity}
\bea
\langle T_{ij}\rangle&=&-\frac{2}{\sqrt{h_{(0)}}}\frac{\delta S_{ren}}{\delta h_{(0)}^{ij}}\nonumber\\
&=&\lim_{\epsilon\rightarrow0}\frac{T_{ij}[\gamma]}{\epsilon^2}\nonumber\\
&=&\frac{1}{\kappa^2}\left(K_{ij}-\gamma_{ij}K\right)+\frac{2}{\sqrt{\gamma}}\frac{\delta S_{ct}}{\delta \gamma^{ij}}.
\eea

\section{Weak anisotropy and small charge density  analysis}
\label{highApp}
In the follows, we will  try to find the perturbative solution  around the isotropic black brane solution, that is to say, the anisotropic parameter $a$ will be treated as an expansion parameter in solving  the eqs. (\ref{eq_F_app})-(\ref{3order_dil}). To do so, we expand all the coefficients up to the order $a^4$ as followed:
\bea
\cf(u)&=& \cf_0(u) + a^2  \cf_2 (u)+ a^4  \cf_4(u) + {O}(a^6) \,,  \\[1.7mm]
\cb(u) &=& \cb_0(u)+ a^2 \hat \cb_2 (u)+ a^4   \cb_4(u) + {O}(a^6) \,, \\[1.7mm]
\phi(u) &=& a^2  \phi_2 (u)+ a^4  \phi_4(u) + {O}(a^6) .
\label{small_a_exp}
\eea
 Note that due to the symmetry $z\to -z$, only even orders appear, and the absence of 0-order expansion of dilaton is attributed to the early statement that $\phi$ is vanishing when $a=0$.

Now what we have to do next is to determine the coefficient functions $ \cf_n(u)$, $ \cb_n(u)$, $ \phi_n(u)$. We then substitute these expansions into Einstein's equations and solve them order by order in $a$.
The asymptotically Anti-de Sitter condition requires that $\cf(u)=\cb(u)=1$ and $\phi(u)=0$ at the boundary $u=0$. Additionally, by definition for horizon, $\cf(u)$ vanishes at the horizon
which means that   $\cf_n(u)=0$ at the horizon $u=\uh$. In this appendix, we  focus on solutions with in the small $q$ limit so that they are in the high temperature regime.
%%%%%%%%%%%%
\subsection{0-order}
We start with solving 0-order term of the coefficient equations. Since we require 0-order of $\phi$ to be vanished, what we have to solve is $\cf_0$ and $\cb_0$, which satisfy the boundary conditions
\bea
\cf_0(0)=1 , ~~~~\cb_0(0)=1.
\eea
The Einstein equations can be solved directly with  a simple and familiar form
 \bea
 \cf_0(u)&=&1-(\frac{u}{\uh})^4+\Big[(\frac{u}{\uh})^6-(\frac{u}{\uh})^4\Big]q^2\,,\nonumber\\[1.7mm]
 \cb_0(u)&=&1\,,
 \eea
 where the $q$ is a dimensionless charge parameter  with $q=\frac{u_H^3 Q}{2\sqrt{3}}$. So the 0-order solution is nothing but the RN-AdS black brane as we expected.
\subsection{$O(a^2)$ order}\label{c1}

To the 2nd-order of $a$, we can obtain the solutions in the small $q$ limit. So we can expand the coefficients $\cf_2(u), \cb_2(u), \phi_2(u)$  as follows
\bea
\cf_2(u)&=&\hat{\cf_0}(u)+\hat{\cf_2}(u) q^2+{O}(q^4),   \nonumber \\[0.7mm]
\cb_2(u)&=&\hat{\cb_0}(u)+\hat{\cb_2}(u) q^2+{O}(q^4) ,   \nonumber \\[0.7mm]
\phi_2(u)&=&\hat{\phi_0}(u)+\hat{\phi_2}(u)q^2+{O}(q^4),
\eea
with boundary conditions $\cf_2(0)=\cb_2(0)=0$, which yield the boundary conditions for the expanded coefficients
\bea
 \hat{\cf}_{2n}(0)&=&0, ~\hat{\cf}_{2n}(\uh)=0, \,  \nonumber\\
 \hat{\cb}_{2n}(0)&=&0, \nonumber\\
 \hat{\phi}_{2n}(0)&=&0 , \, ~~~~(n=0,1,2,...).
 \eea
Plunging these   expressions into the Einstein equations,  we obtain the functions to the second order of $q$:
\bea
\hat\cf_0(u)&=&\frac{1}{24 \uh^2}\left[8 u^2( \uh^2-u^2)-10 u^4\log 2 +(3 \uh^4+7u^4)\log\left(1+\frac{u^2}{\uh^2}\right)\right], \nonumber \\[0.7mm]
\hat\cb_0(u) &=& -\frac{\uh^2}{24}\left[\frac{10 u^2}{\uh^2+u^2} +\log\left(1+\frac{u^2}{\uh^2}\right)\right],                 \nonumber \\[0.7mm]
\hat\phi_0(u)&=& -\frac{\uh^2}{4}\log\left(1+\frac{u^2}{\uh^2}\right)\,,
\eea
and
\bea
\hat\cf_2(u)&=&\frac{1}{24 \uh^4(u^2+\uh^2)} \Big[7u^8+6u^2\uh^6+u^4\uh^4(25\log2-12)\nonumber \\
&&~~~~~~~~~~+u^6\uh^2(25\log2-1)-(u^2+\uh^2)(12u^6+7u^4\uh^2+6\uh^6)\log\left(1+\frac{u^2}{\uh^2}\right) \Big], \nonumber \\[0.7mm]
\hat\cb_2(u) &=& \frac{1}{24}\Big[-\frac{u^2(11u^4+3u^2\uh^2+2\uh^4)}{(u^2+\uh^2)^2}+2\uh^2\log\left(1+\frac{u^2}{\uh^2}\right)\Big],                    \nonumber \\[1.7mm]
\hat\phi_2(u)&=& \frac{1}{4}\Big[-2u^2+\frac{u^4}{u^2+\uh^2}+2\uh^2\log\left(1+\frac{u^2}{\uh^2}\right)\Big].
\eea
So, using(\ref{potential}), we can read the electric potential $A_t$ as
\bea
A_t=\frac{q}{8\sqrt{3}\uh^3}\left[24(\uh^2-u^2)+5a^2\uh^2\left(u^2\log(1+\frac{u^2}{\uh^2})-\uh^2\log2\right)\right]
\eea
It is straightforward to check that (\ref{max exp}) holds after coordinates transformation. Then the charge density $\rho$ and the corresponding chemical potential $\mu$ can be extracted from the asymptotic expansion of Maxwell filed (\ref{Max exp}), one obtain the results
\bea
\mu=\frac{q(24-5a^2\uh^2\log2)}{8\sqrt{3}\uh},~ ~~~~ \rho=\frac{\sqrt{3}q}{\kappa^2\uh^3}.
\eea
We can immediately  obtain the temperature of the system to order $a^2$ by evaluating these expressions at the horizon:
\bea
T= \frac{2-q^2}{2\pi\uh}+\frac{\uh[10\log2-4+5(3+\log2)q^2]}{96\pi}a^2+{O}(a^4)\,.\label{tem2}
\eea
 Using (\ref{tem2}), we see that the entropy density can be expressed as
\bea
s=\frac{\nc^2 e^{-\frac{5}{4}\phi_\textrm{H}}}{2\pi \uh^3}=\frac{\pi^2T^3\nc^2}{2}+\frac{3\pi^2\nc^2T^3}{4}q^2+\left(\frac{\nc^2T}{16}+\frac{(2-15\log2)\nc^2T}{32}q^2\right)a^2+{O}(a^4). \nonumber\\
\eea
So the heat capacity with fixed charge of the black brane is
\bea
c_{\rho,a} =T(\frac{\partial s}{\partial T})_{\rho,a}
=\frac{3\pi^2\nc^2 T^3}{2}+\frac{9\pi^2\nc^2 T^3}{4}q^2+ \frac{(2 + q^2 (2 -15\log2))T\nc^2}{32} a^2+{O}(a^4). \nonumber\\
 \eea

  After obtaining the high temperature solution, we now  evaluate the energy and pressures from the asymptotic expansions of the fields.
Note that  $\mathbb{F}_4$ and $\mathbb{B}_4$ take the form
\bea
\mathbb{F}_4&=&-\pi^4T^4-3\pi^4T^4q^2+\left(-\frac{9}{16}\pi^2T^2+\frac{1}{16}\pi^2T^2(-13+30\log2)q^2\right)a^2+{O}(a^4). \nonumber\\
\mathbb{B}_4&=&\left(\frac{7\pi^2T^2}{16}+\frac{7\pi^2T^2q^2}{16}\right)a^2+{O}(a^4).
\eea
Substituting these expressions into (\ref{stress_tensor}), we obtain the energy density and the pressures as
\bea
E&=&\frac{3\pi^2\nc^2T^4}{8}+\frac{9\pi^2\nc^2T^4}{8}q^2+\left(\frac{\nc^2T^2}{32}+\frac{(8-45\log2)\nc^2T^2q^2}{64}\right)a^2+{O}(a^4),\nonumber \\
P_{xy}&=&\frac{\pi^2\nc^2T^4}{8}+\frac{3\pi^2\nc^2T^4}{8}q^2+\left(\frac{\nc^2T^2}{32}+\frac{(4-15\log2)\nc^2T^2}{64}q^2\right)a^2+{O}(a^4),\nonumber \\
P_z&=&\frac{\pi^2\nc^2T^4}{8}+\frac{3\pi^2\nc^2T^4}{8}q^2+\left(-\frac{\nc^2T^2}{32}-\frac{(15\log2)\nc^2T^2}{64}q^2\right)a^2+{O}(a^4).
\label{2order-stress}
\eea
It is easy to see that the conformal anomaly $\langle T_\mu^\mu\rangle$ vanishes.

Now we can see that the thermodynamical relations in section \ref{sec-thermodynamics} are reasonable. First of all, the Free energy density given by (\ref{Free energy}) when evaluating our specific solution reads
\bea
\Omega=\left(-\frac{\nc^2 \pi ^2 T^4}{8} -\frac{3\pi ^2\nc^2 T^4}{8}q^2 \right)+\left(-\frac{\nc^2T^2}{32}+\frac{(15\log2-4)\nc^2T^2q^2}{64}\right)a^2+{O}(a^4).
\eea
It is easily to prove that
\bea
\Omega&=&E-Ts-\rho\mu.
\eea
Then the ``chemical potential" corresponding to the axion filed is expressed as
\bea
\Phi = \left( \frac{\partial \Omega}{\partial a} \right)_T = -\frac{\nc^2T^2}{16}a-\frac{\nc^2T^2q^2}{16}a+{O}(a^3).\,
\eea
Note that in the calculation above, $\uh$ should be thought as a function of $a$ .
\subsection{$O(a^4)$ order}\label{lo}
Similar argument  leads to the solutions for Einstein equations to the $O(a^4)$ order. Since the full expressions is very cumbersome, we just list the asymptotic expansions near the boundary
\bea
\cf_4(u)&=&\frac{1}{3456}\left(-915-40\pi^2+1611 \log 2+1440(\log 2)^2+2016 \log\frac{u}{\uh}\right)u^4 \nonumber \\
&+&\frac{q^2}{2304}\big(699+160\pi^2-6\log2\left(36+785\log2\right)\big)u^4+{O}(u^6), \nonumber\\
\cb_4(u)&=&  \frac{1}{1152}\left(551-567 \log 2-672 \log\frac{u}{\uh}\right)u^4+\frac{7q^2}{768} (-49 + 76 \log2)u^4+{O}(u^6), \nonumber\\
\phi_4(u)&=& \frac{1}{192}\left(32-27 \log 2-32 \log\frac{u}{\uh}\right)u^4+\frac{q^2}{384}(-49 + 76 \log2)u^4+{O}(u^6).
\eea
To compute the temperature and the entropy density, we also need the coefficients evaluated on the horizon
\bea
\cf_4'(\uh)&=&\frac{\uh^3\left[6-4\pi^2+(42+165\log2)\log2+(42+14\pi^2-138\log2(-1+2\log2))q^2\right]}{288},\nonumber\\
\cb_4(\uh)&=&\frac{\uh^4\left[369-8\pi^2+354(\log2)^2+912\log2+(411+44\pi^2-1488(\log2)^2+1386\log2)q^2\right]}{6912},\nonumber\\
\phi_4(\uh)&=&\frac{\uh^4\left[6-4\pi^2+174(\log2)^2-12\log2+\left(-51+22\pi^2+3(99-244\log2)\log2\right)q^2\right]}{576}.
\eea
So the temperature and the entropy density are given by
\bea
T&=&\frac{2-q^2}{2\pi\uh}+\frac{\uh\left[10\log2-4+5(3+\log2)q^2\right]}{96\pi}a^2+\Big( \frac{[40\pi^2+3(44+15(4-37\log2)\log2)]\uh^3}{13824\pi}\nonumber\\
&-&\frac{\left[468+80\pi^2+(558-1155\log2)\log2\right]\uh^3}{9216\pi}q^2\Big)a^4+{O}(a^6),~~~~~\\
s&=&\frac{\nc^2}{2\pi\uh^3}+\left(\frac{10\log2+(15-20\log2)q^2}{64\pi\uh}\nc^2\right)a^2\nonumber\\
&+&\frac{5\uh(8\pi^2-12+24\log2-303(\log2)^2-(44\pi^2-102+459\log2-1284(\log2)^2)q^2)}{9612\pi}a^4+{O}(a^6) ,\nonumber\\
\eea
respectively. Again, the chemical potential corresponding to the $U(1)$ gauge filed is found to be
\bea
\mu=\frac{q(24-5a^2\uh^2\log2)}{8\sqrt{3}\uh}+\frac{q\uh^3\left(222-20\pi^2+(-348+945\log2)\log2\right)}{1152\sqrt3}a^4+{O}(a^6).
\eea
The expressions for $\mathbb{F}_4$ and $\mathbb{B}_4$ are obtained as
\bea
\mathbb{F}_4=-\frac{1+q^2}{\uh^4}&+&\left(-\frac{19+20\log2-(30-50\log2)q^2}{48\uh^2}\right)a^2\nonumber\\
&&~~~~~+\bigg(\frac{-915-40\pi^2+(1611\log2+1440(\log2)^2-2016\log\uh)}{3456}\nonumber\\
&&~~~~~~~~~~~~+\frac{699+160\pi^2-216\log2-4710(\log2)^2}{2304}\bigg)a^4+{O}(a^6),\nonumber\\
\mathbb{B}_4=\frac{7}{16\uh^2}a^2&+&\frac{1}{2304}\Big(1102-1134\log2+21q^2(-49+76\log2)+1344\log\uh\Big)a^4+{O}(a^6),~~~~~~~
\eea
which follow that the energy and pressures are given by
\bea
E&=&\frac{3+3q^2}{8\pi^2\uh^4}\nc^2+\left(\frac{10\log2-2+(15-25\log2)q^2}{64\pi^2\uh^2}\right)a^2\nonumber\\
&+&\Big(\frac{(24\cs+20\pi^2-51+18(7-40\log2)\log2-96\log\uh)\nc^2}{4608\pi^2}\nonumber\\&+&\frac{(428-160\pi^2+(-1532\log2+4710(\log2)^2))\nc^2}{6144\pi^2}q^2\Big)a^4+{O}(a^6),\nonumber\\[1.7mm]
P_{xy}&=&\frac{1+q^2}{8\pi^2\uh^4}\nc^2+\left(\frac{10\log2+2+(15-25\log2)q^2}{192\pi^2\uh^2}\right)a^2\nonumber\\
&+&\Big(\frac{(-72\cs+20\pi^2+129-198\log2-720(\log2)^2+288\log\uh)\nc^2}{4608\pi^2}\nonumber\\&+&\frac{(-80\pi^2+3(6+\log2(-154+785\log2)))\nc^2}{9216\pi^2}q^2\Big)a^4+{O}(a^6),\nonumber\\[1.7mm]
P_{z}&=&\frac{1+q^2}{8\pi^2\uh^4}\nc^2+\left(\frac{10\log2-10+(15-25\log2)q^2}{192\pi^2\uh^2}\right)a^2\nonumber\\
&+&\Big(\frac{(1512\cs+140\pi^2+789+126(43-40\log2)\log2-6048\log\uh)\nc^2}{96768\pi^2}\nonumber\\&+&\frac{(-80\pi^2+3(202+\log2(-458+785\log2)))\nc^2}{9216\pi^2}q^2\Big)a^4+{O}(a^6).
\label{4order-stress}
\eea
Thus, the conformal anomaly is $\langle T_i^i\rangle=\nc^2 a^4/48\pi^2$. This fact implies that the $O(a^4)$ term is the reflection of the quantum effect because the conformal anomaly is finite once we take account into the $a^4$ term. Note that we have not included the reference scale $\Lambda$ required by the dimension analysis and the presence of the conformal anomaly, but we shall add this $\log$ term (see the  discussion in Section 3).

The thermodynamic potential is given by
\bea
\Omega=-\frac{1+q^2}{8\pi^2\uh^4}\nc^2&-&\left(\frac{10\log2+2+(15-25\log2)q^2}{192\pi^2\uh^2}\nc^2\right)a^2\nonumber\\
&-&\Big(\frac{(-72\cs+20\pi^2+129-198\log2-720(\log2)^2+288\log\uh)\nc^2}{4608\pi^2}\nonumber\\&+&\frac{(-80\pi^2+3(6+\log2(-154+785\log2)))\nc^2}{9216\pi^2}q^2\Big)a^4+{O}(a^6),
\eea
from which we obtain the chemical potential associated with the axion
\bea
\Phi&=&-\frac{\nc^2a}{16\pi^2\uh^2}+\Big(\frac{\nc^2(-19+336\cs+1134\log2-1344\log\uh)}{16128\pi^2}\nonumber\\
&+&\frac{\nc^2(1029-1596\log2)q^2}{16128\pi^2}\Big)a^3+{O}(a^5).
\eea
\section{Weak anisotropy and finite charge density analysis}
\label{low App}
In the previous section, we  obtained the metric solution by perturbing around the Schwarzschild black brane solution in the small $a$ and small $q$ limit. It is straightforward to obtain the solution by perturbing around the RN-AdS black brane solution. In fact for finite charge, we can only obtain the analytic expression up to $\mathcal{O}(a^2)$ order.
 The functions $\cf$, $\cb$ and $\ch$ can be expressed as
 \bea
 &&\cf=1-\bigg(\frac{u}{u_H}\bigg)^4+\bigg[\bigg(\frac{u}{u_H}\bigg)^6-\bigg(\frac{u}{u_H}\bigg)^4\bigg]q^2+a^2 \cf_2(u)+\mathcal{O}(a^4),\nonumber\\[1.7mm]
 &&\cb=1+a^2 \cb_2(u)+\mathcal{O}(a^4),\nonumber\\[1.7mm]
 &&\ch=e^{-\phi(u)}, {~~~\rm with ~~~}  \phi(u)=a^2 \phi_2(u)+\mathcal{O}(a^4),
 \eea
 where
 \bea
 \cf_2(u)&=&\frac{1}{24\sqrt{1+4q^2}\uh^4}\bigg\{3(-4q^2u^6+\uh^6)\log\left(\frac{(1+\sqrt{1+4q^2})u^2+2\uh^2}{(1-\sqrt{1+4q^2})u^2+2\uh^2}\right) \nonumber\\[0.7mm]
 &+&u^2\uh^2\Big[8\sqrt{1+4q^2}(-u^2+\uh^2)+u^2\Big(3\log\left(-2-2q^2+2\sqrt{1+4q^2}\right)\nonumber\\[0.7mm]
 &+&5(-2+q^2)\log\left(-1+2q^2+\sqrt{1+4q^2}\right)-12q^2\log\left(-2-2q^2+2\sqrt{1+4q^2}\right)\nonumber\\[0.7mm]
 &+&7(1+q^2)\Big(\log\left((-1+2q^2-\sqrt{1+4q^2})(2q^2u^2+(-1+\sqrt{1+4q^2})\uh^2)\right)\nonumber\\[0.7mm]
 &-&\log\left(2q^2u^2-(1+\sqrt{1+4q^2})\uh^2\right)\Big)\Big)\Big]\bigg\},\nonumber\\[1.7mm]
 \phi_2(u)&=&-\frac{\uh^2}{4\sqrt{1+4q^2}}\log\left(\frac{(1+\sqrt{1+4q^2})u^2+2\uh^2}{(1-\sqrt{1+4q^2})u^2+2\uh^2}\right),\nonumber\\[1.7mm]
 \cb_2(u)&=&\frac{\uh^2}{24}\left(\frac{10u^2\uh^2}{q^2u^4-u^2\uh^2-\uh^4}+\frac{1}{\sqrt{1+4q^2}}\log\left(\frac{(1+\sqrt{1+4q^2})u^2+2\uh^2}{(1-\sqrt{1+4q^2})u^2+2\uh^2}\right)\right).
 \eea
 In terms of the  series expansion of $O(q^2)$, we can recover the result obtained in \ref{c1}.
 The Hawking temperature and entropy density are given by
 \bea\label{tema}
 T=\frac{2-q^2}{2\pi \uh}+\frac{\uh\left(-4\sqrt{1+4q^2}+5(2+5q^2)\log\big(\frac{3+\sqrt{1+4q^2}}{3-\sqrt{1+4q^2}}\big)\right)}{96\pi\sqrt{1+4q^2}}a^2+\mathcal{O}(a^4),
 \eea
and
 \bea
 s=\frac{N^2_c}{2\pi \uh^3}+\frac{5\nc^2\log\big(\frac{3+\sqrt{1+4q^2}}{3-\sqrt{1+4q^2}}\big)}{32\pi\sqrt{1+4q^2}\uh}a^2+\mathcal{O}(a^4).
 \eea
  We notice that for $a>0$ at finite temperature, the horizon radius $\uh$ and the entropy density  of the anisotropic black brane are greater than that of isotropic RN-AdS black brane.
 This partly verifies our previous numerical computation that extremal black brane solution cannot  be accessed for $a>0$ .

 \begin{figure}[htbp]
 \begin{minipage}{1\hsize}
\begin{center}
%\vspace*{10mm}
\includegraphics*[scale=0.5] {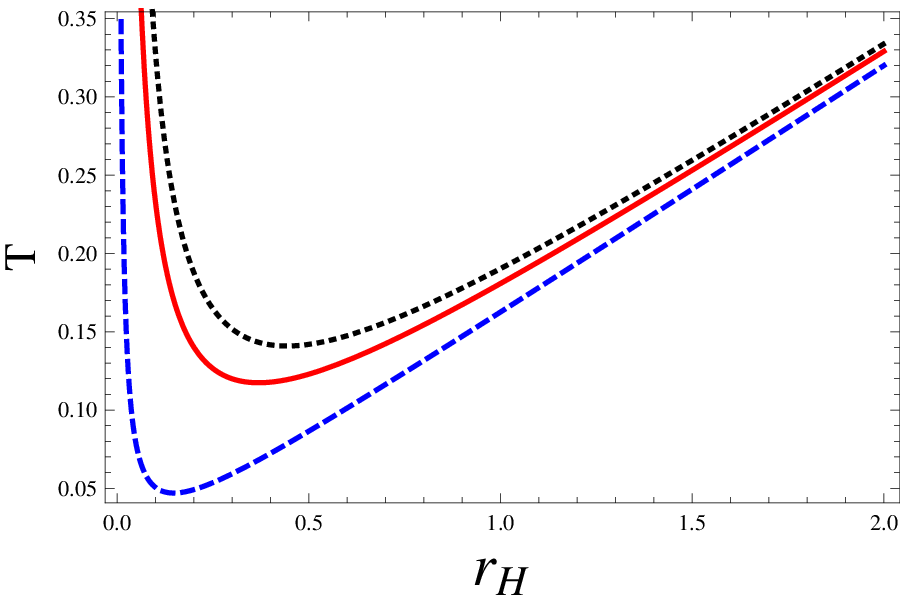}
\includegraphics*[scale=0.5] {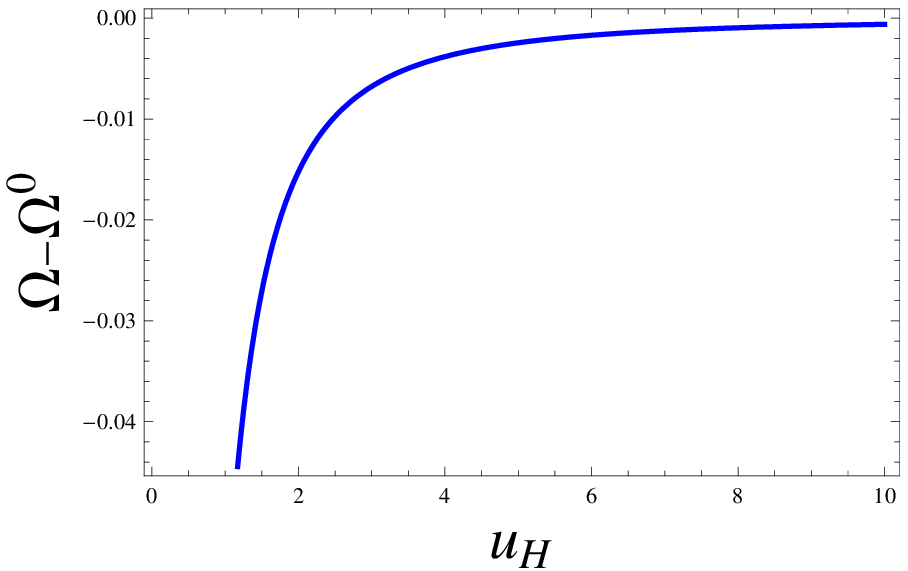}
\includegraphics*[scale=0.5] {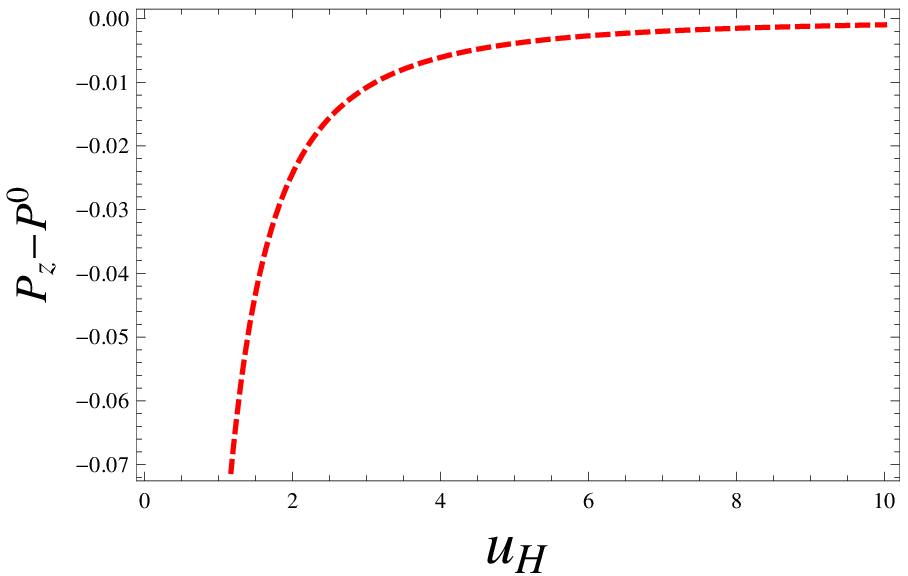}
\end{center}
\caption{(Color online.)(Left) Temperature as a function of the horizon radius for $a=1/5$(blue), $a=1/2$(red) and $a=3/5$ (black), where we choose $N_c=1$ and $q^2=1$. (Middle) We see that the anisotropic black brane has lower grand thermodynamic potential with respect to the isotropic RN-AdS solution. (Right) The plot shows that the pressure along the $z$-direction is smaller than that of the isotropic solution. In all the three graphs, we do not consider the contribution of the
$\log(a/\Lambda)$ term.} \label{solx}
\end{minipage}
\end{figure}
 The energy density and pressures are obtained as follows
\bea
 E&&=\frac{3(1+q^2)\nc^2}{8\pi^2\uh^4}+\frac{\nc^2\Big(-2\sqrt{1+4q^2}+5(q^2-2)\log\left(\frac{3-\sqrt{1+4q^2}}{3+\sqrt{1+4q^2}}\right)\Big)}{64\pi^2\sqrt{1+4q^2}\uh^2}a^2+{O}(a^4),\nonumber\\[0.3mm]
 P_{xy}&&=\frac{(1+q^2)\nc^2}{8\pi^2\uh^4}+\frac{\nc^2\Big(2\sqrt{1+4q^2}+5(q^2-2)\log\left(\frac{3-\sqrt{1+4q^2}}{3+\sqrt{1+4q^2}}\right)\Big)}{192\pi^2\sqrt{1+4q^2}\uh^2}a^2+{O}(a^4),
 \nonumber\\
 P_z&&=\frac{(1+q^2)\nc^2}{8\pi^2\uh^4}+\frac{5\nc^2\Big(-2\sqrt{1+4q^2}+(q^2-2)\log\left(\frac{3-\sqrt{1+4q^2}}{3+\sqrt{1+4q^2}}\right)\Big)}{192\pi^2\sqrt{1+4q^2}\uh^2}a^2+{O}(a^4).
 \eea
The ``chemical" potential conjugate to the charge ``a" is given by
\bea
\Phi=-\frac{ \nc^2 a}{16 \pi^2 \uh^2}+{O}(a^3).
\eea
The $U(1)$ chemical potential is obtained as
\bea
\mu&&=\frac{q}{8\sqrt{3}\uh}\left(24+\frac{5\uh^2\log(\frac{3-\sqrt{1+4q^2}}{3+\sqrt{1+4q^2}})}{\sqrt{1+4q^2}}a^2\right)+{O}(a^4).
 \eea
 Unfortunately, we are not able to obtain the analytic solution up to $\mathcal {O}(a^4)$.
 Fig.\ref{solx} (left) shows us that the temperature has its minimize value as a function of the horizon radius for given $a$ and $q$. Moreover, in the absence of the reference scale $\Lambda$,
 the thermodynamic potential and the pressure $P_z$ are lower than that of the isotropic case.

 Considering the presence of the conformal anomaly and the dimensional analysis, we shall include the reference scale $\Lambda$ to the energy stress tensor $E$, $P_{xy}$, $P_z$ and $\Phi$, even in the absence of
 the $\mathcal {O}(a^4)$ (see discussion in Section \ref{phase}).

\section{The ``oblate" solution}
\label{imag}
In this appendix, we  take a brief look at the five-dimensional anisotropic black brane solution in the case of imaginary-valued $a$. As claimed in the main text, imaginary $a$ could lead to the zero temperature, i.e. extremal anisotropic black brane. However, in this case, it is not proper to interpret the anisotropy $a$  as the D7-brane number or ``charge density" (\ref{number}) in the framework of string/gauge duality. We can only regard $a$  as a parameter to support the anisotropic black brane geometry, just like some gauge fields in charged Lifshitz theory are only treated as extra matter fields to accommodate a Lifshitz spacetime without a clear signature in the thermodynamics of the system \cite{lifshitz}.

\begin{figure}[htbp]
 \begin{minipage}{1\hsize}
\begin{center}
%\vspace*{10mm}
\includegraphics*[scale=0.50] {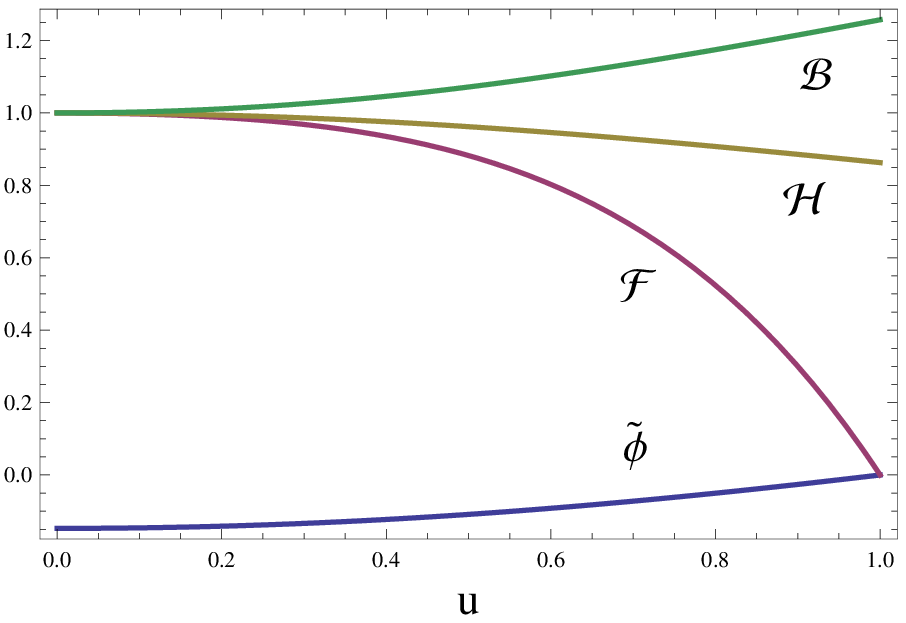}
\includegraphics*[scale=0.50]{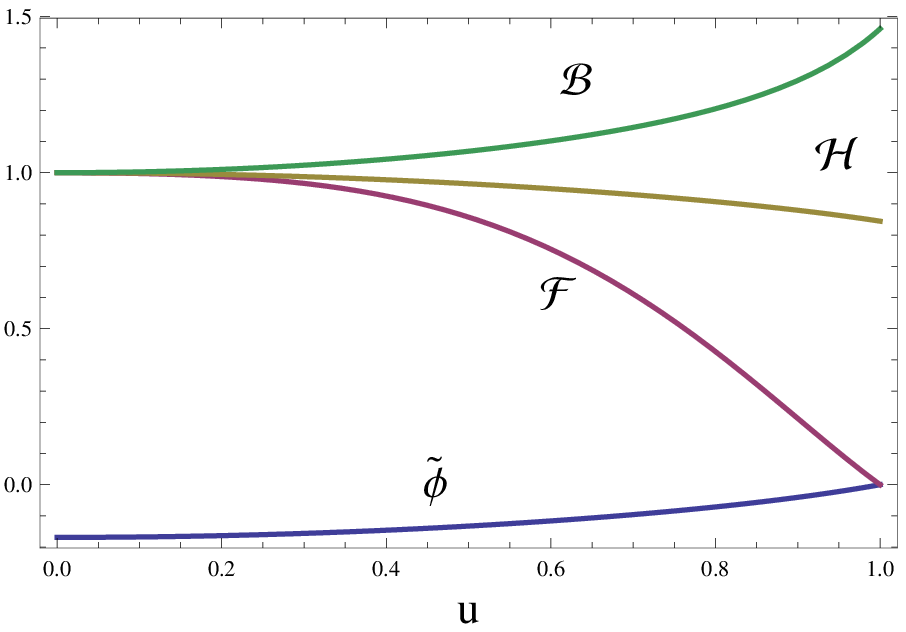}
\includegraphics*[scale=0.52] {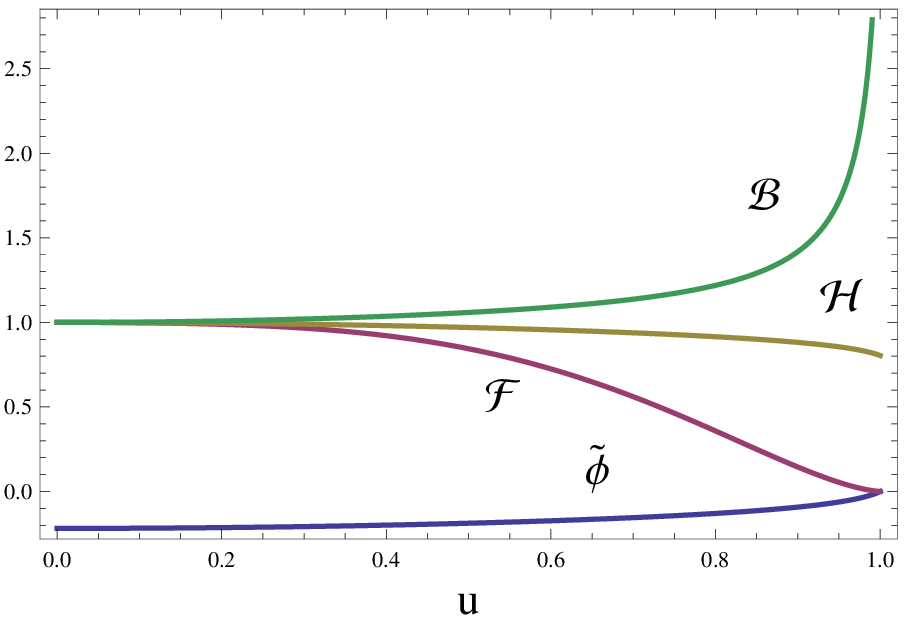}
\end{center}
\caption{(color online)The metric functions for $\tilde{Q}=0.25$ (left),  $\tilde{Q}=10$ (middle) and $\tilde{Q}=20$ (right), with $\tilde\phi_H=0$ and $u_H=1$.} \label{neiga1}
\end{minipage}
\end{figure}

\begin{figure}[htbp]
 \begin{minipage}{1\hsize}
\begin{center}
%\vspace*{10mm}
\includegraphics*[scale=0.50] {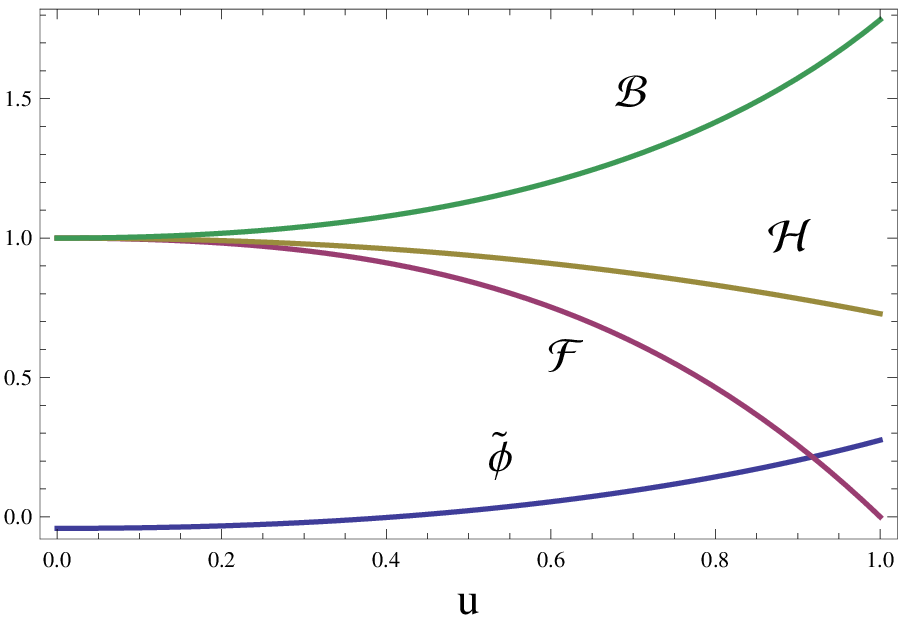}
\includegraphics*[scale=0.50]{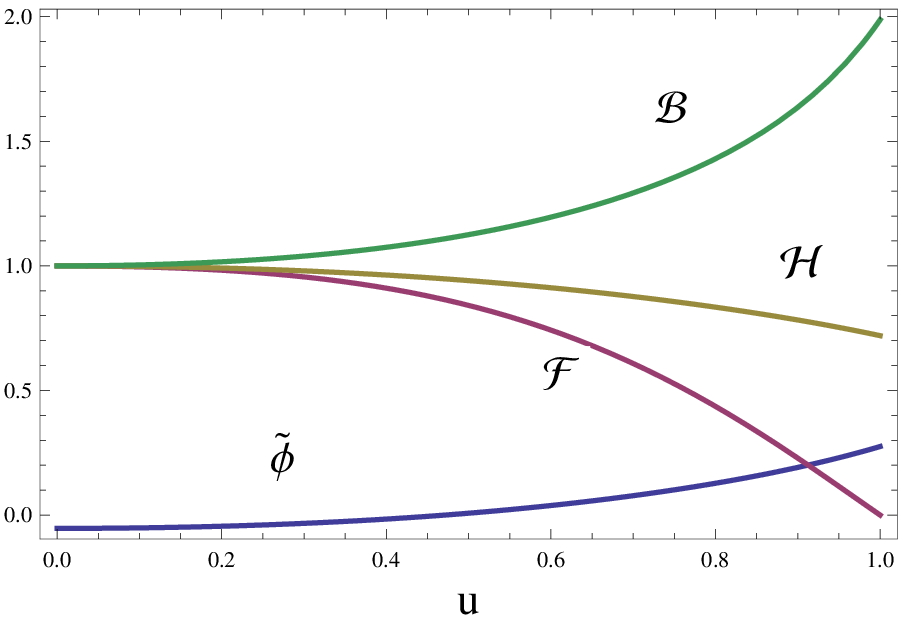}
\includegraphics*[scale=0.52] {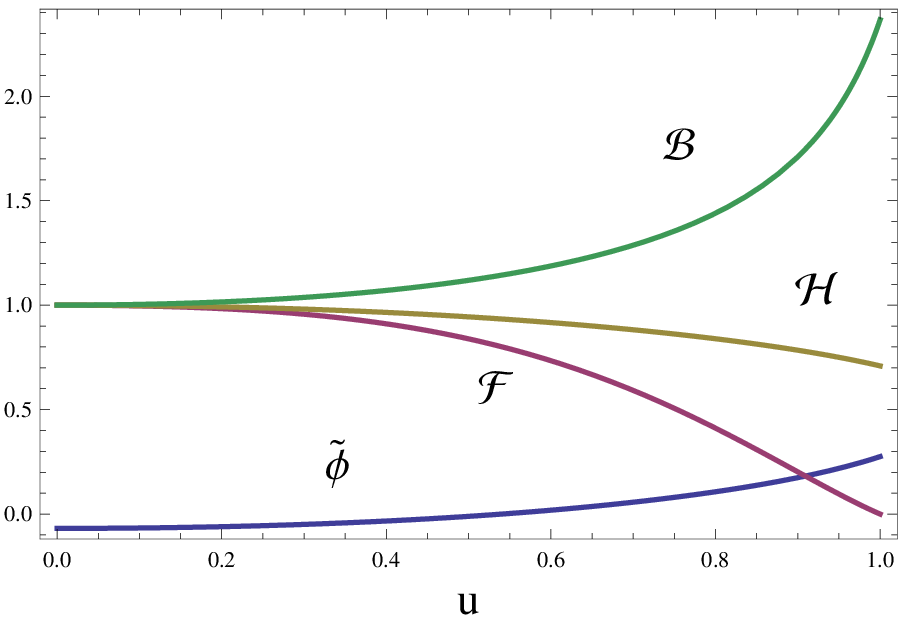}
\end{center}
\caption{(color online)The metric functions for $\tilde{Q}=0.1$ (left),  $\tilde{Q}=2$ (middle) and $\tilde{Q}=4$ (right), with $\tilde\phi_H=0.275$ and $u_H=1$.} \label{neiga2}
\end{minipage}
\end{figure}

\begin{figure}[htbp]
 \begin{minipage}{1\hsize}
\begin{center}
%\vspace*{10mm}
\includegraphics*[scale=0.52] {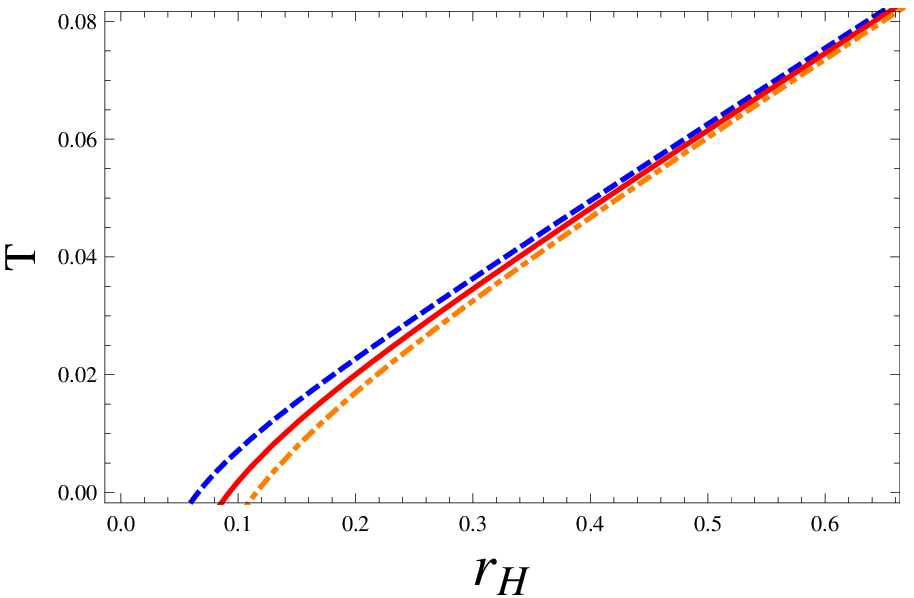}
\includegraphics*[scale=0.54] {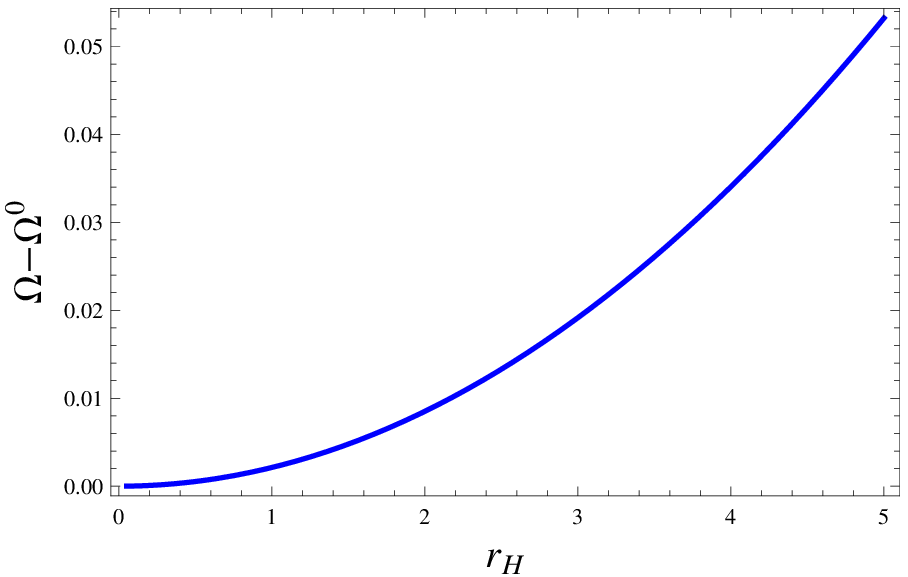}
\includegraphics*[scale=0.55] {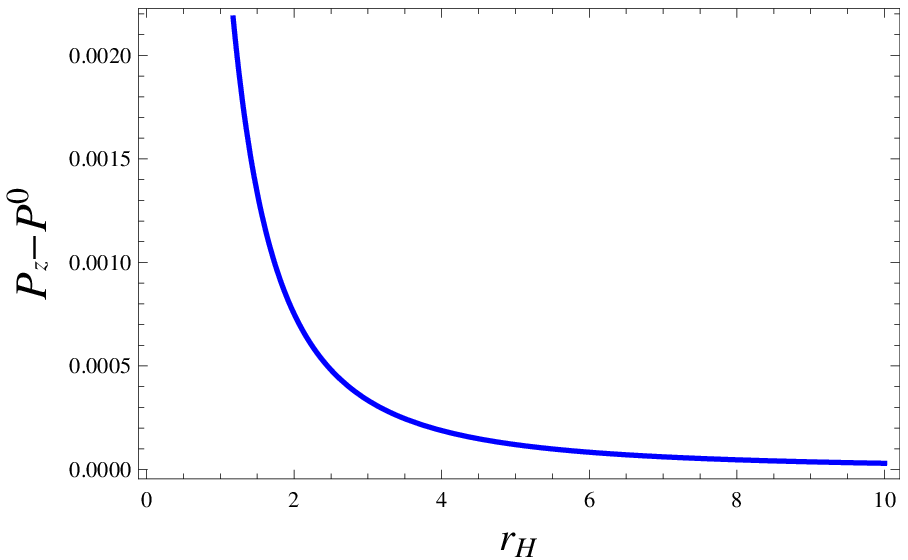}
\end{center}
\caption{(Left) Temperature as a function of the horizon radius $r_{H}=1/u_{H}$ for $a=i/14$(blue), $a=i/10$(red) and $a=i/8$ (orange), where we choose $N_c=1$ and $q^2=1.2$. (Middle) The free energy
as a function of the horizon radius $r_{H}$, where $N_c=8$, $\Lambda=1/2$ and $q^2=1/4$ .
  The center graph shows that  anisotropic black brane has higher grand thermodynamical potential compared to the isotropic RN-AdS solution. On the right, the graph depicts  that the pressure along the $z$-direction is greater than that of the isotropic solution. Also, we do not consider the contribution of the
$\log(a/\Lambda)$ term.}\label{solu}
\end{minipage}
\end{figure}

We notice that the concrete form of equations of motion is independent of the exact value of $a$. So we can find the solutions numerically  or analytically in a similar way. Since $a^2<0$, the shift of the  dilaton should take the form of
\bea
\tilde{\phi}=\phi+\frac{2}{7}\log(-a^2).
\eea
The boundary condition of the dilaton  $\phi(0)=0$ gives
\bea
a^2=-e^{\tilde{\phi}(0)}.
\eea
and the boundary conditions given in (\ref{bcondition}) should be shifted as $e^{7\tilde\phi_H/2}\rightarrow-e^{7\tilde\phi_H/2}$.  We can also introduce a charge related parameter $\tilde{Q}=(-a^2)^{5/7} Q^2$. Then we can plot the numerical solutions in Fig\ref{neiga1} and Fig\ref{neiga2}.
As shown in the numerical solutions, the anisotropic factor $\ch(u)=e^{-\phi(u)} \leq1$, which implies that this solution is an``oblate" black brane whose  z-direction is shorter than the $x-$ and $y-$ axes. Different from the ``prolate" case, the profile for $\cb$ here is amplified at the horizon as $\tilde{Q}$ increases.
The  ``oblate" black brane has a significant property: the temperature can be zero as shown Fig \ref{solu}.

On the analytic side, if we treat axion as small anisotropic fluctuation, the analytic solution can also be obtained which is the same with previous section. A straightforward calculation demonstrates that the positiveness of the horizon radius $u_H$ requires that the axion parameter $a$ must be imaginary when the horizon radius $\uh$ takes the form
\bea \label{zeroT}
\uh={4\sqrt{3}[(2-q^2)\sqrt{1+4q^2}]^{1/2}}{\bigg[a^2\bigg(4\sqrt{1+4q^2}-5(2+5q^2)\log\big(\frac{3+\sqrt{1+4q^2}}{3-\sqrt{1+4q^2}}\big)\bigg)\bigg]^{-1/2}}.
\eea
In the  case that $a=0$, the black brane solution recovers the extremal RN-AdS case with zero temperature but non-vanishing entropy.
If the parameter $a$ takes imaginary value, the thermodynamic variables of this system becomes totally different. In this case, for a given temperature $T$, the horizon radii $r_{H}\equiv\frac{1}{\uh}$ and the entropy density are less than those of  RN-AdS black brane. We also note that $\mu>\mu^0$ in this case.
Fig.\ref{solu} (left) plots the temperature as a function of the horizon radius. The temperature becomes a monotonic function of the horizon radius and zero temperature is available for imaginary $a$. We can also see from Fig.\ref{solu} (right) that  oblate black brane has higher grand thermodynamical potential than the isotropic RN-AdS solution..
%%%%%%%%%%%%%%%%%%%%%%%%%%%

We end this appendix with some comments. For a scalar field $\phi$ with mass $m$ in asymptotic $AdS_5$ space, the well-known Breitenlohner-Freedman (BF) bound \cite{BF} on the stability condition of the background requires $m^2\geq -4$. In our case, in the dilatonic equation of motion $\nabla_\mu\nabla^\mu\phi-e^{2\phi}(\partial\chi)^2=0$,
$e^{2\phi}(\partial\chi)^2$ corresponds to  the mass term.  It is reasonable to speculate that if the imaginary value of $a$ takes some value, the dilaton field becomes tachyonic. So that, the dilaton could condensate in the IR, similar to that observed in the condensation of neutral scalar fields in holographic superconductors.

\end{document}